\documentclass{sigplanconf}

\usepackage{mathpartir}
\usepackage{amsmath}
\usepackage{amsfonts}
\usepackage{comment}
\usepackage{flushend}
\usepackage{stmaryrd}
\usepackage{xspace}
\usepackage[pdftex,
            pdfauthor={Sam Lindley, Conor McBride, and Craig McLauglin},
            pdftitle={Do be do be do}]{hyperref}
\hypersetup{colorlinks=true,allcolors=black}
\usepackage[usenames,dvipsnames]{xcolor}
\usepackage{url}

\usepackage{MnSymbol}

\usepackage{etoolbox}

\makeatletter
\pretocmd{\NAT@citexnum}{\@ifnum{\NAT@ctype>\z@}{\let\NAT@hyper@\relax}{}}{}{}
\makeatother

\newcommand{\fighead}{\textbf}

\newcommand{\lameff}{$\lambda_{\text{eff}}$\xspace}

\newcommand{\EC}{\mathcal{E}}

\newcommand{\many}{\overline}

\newcommand{\sem}[1]{\llbracket{#1}\rrbracket}

\newcommand{\reducesto}{\longrightarrow}

\newcommand\ba{\begin{array}}
\newcommand\ea{\end{array}}

\newcommand{\bl}{\ba[t]{@{}l@{}}}
\newcommand{\el}{\ea}

\newcommand{\bstack}{\begin{array}[t]{@{}l@{}}}
\newcommand{\estack}{\end{array}}

\newenvironment{equations}{\[\ba{@{}r@{~}c@{~}l@{}}}{\ea\]\ignorespacesafterend}
\newenvironment{eqs}{\ba{@{}r@{~}c@{~}l@{}}}{\ea}

\newenvironment{syntax}{\[\ba{@{}l@{~}r@{~}c@{~}l@{}}}{\ea\]\ignorespacesafterend}

\newcommand{\sigentails}[1]{\mathbin{[{\text{\scriptsize ${#1}$}}]\!\text{-\!-}}\,}

\newcommand{\inferbase}[4]{#2 \mathbin{#1} {#3} \Rightarrow {#4}}
\newcommand{\checkbase}[4]{#2 \mathbin{#1} #4 \mathbin{:} #3}
\newcommand{\patbase}[4]{{#3} \mathbin{:} {#2} \mathbin{#1} {#4}}

\newcommand{\makes}[4]{\inferbase{\sigentails{#2}}{#1}{#3}{#4}}
\newcommand{\has}[4]{\checkbase{\sigentails{#2}}{#1}{#3}{#4}}
\newcommand{\does}[3]{\checkbase{\vdash}{#1}{#2}{#3}}

\newcommand{\makesgs}{\makes{\Gamma}{\sigs}}
\newcommand{\hasgs}{\has{\Gamma}{\sigs}}

\newcommand{\doesg}{\does{\Gamma}}

\newcommand{\infers}{\makes}
\newcommand{\checks}{\has}
\newcommand{\checksdef}{\does}
\newcommand{\matchesc}{\matches}

\newcommand{\infersgs}{\makesgs}
\newcommand{\checksgs}{\hasgs}
\newcommand{\checksdefg}{\doesg}

\newtheorem{theorem}{Theorem}

\newtheorem{definition}[theorem]{Definition}

\newcommand{\sig}{I}
\newcommand{\sigs}{\Sigma}

\newcommand{\effbox}[1]{[#1]}

\newcommand{\key}[1]{\mathbf{#1}} 
\newcommand{\var}{\mathit}        

\newcommand{\op}{\mathsf}  
\newcommand{\con}{\mathsf} 
\newcommand{\inter}{\mathsf} 
\newcommand{\str}[1]{\textrm{``#1''}} 

\newcommand{\handleSymbol}{\rightarrow}
\newcommand{\handle}[2]{{#1} \handleSymbol {#2}}

\newcommand{\thunk}[1]{\{{#1}\}}

\newcommand\slab[1]{(\textrm{#1})}

\newcommand{\adj}{\Delta}

\newcommand{\ev}{E}
\newcommand{\evd}{\varepsilon}

\newcommand{\effin}[1]{\langle {#1} \rangle}
\newcommand{\effout}[1]{[{#1}]}

\newcommand{\nowt}{\emptyset}
\newcommand{\id}{\iota}

\newcommand{\sigyields}[1]{\mathbin{\text{-\!-\!}[{\text{\scriptsize ${#1}$}}]\,}}

\newcommand{\matches}[4]{\patbase{\sigyields{#3}}{#1}{#2}{#4}}
\newcommand{\matchesv}[3]{\patbase{\dashv}{#1}{#2}{#3}}

\begin{document}

\toappear{}



\title{Do Be Do Be Do}


\authorinfo{Sam Lindley}
           {The University of Edinburgh, UK}
           {sam.lindley@ed.ac.uk}
\authorinfo{Conor McBride}
           {University of Strathclyde, UK}
           {conor.mcbride@strath.ac.uk}
\authorinfo{Craig McLaughlin}
           {The University of Edinburgh, UK}
           {craig.mclaughlin@ed.ac.uk}

\maketitle



\begin{abstract}
We explore the design and implementation of Frank, a strict functional
programming language with a bidirectional effect type system designed
from the ground up around a novel variant of Plotkin and Pretnar's
effect handler abstraction.

Effect handlers provide an abstraction for modular effectful
programming: a handler acts as an interpreter for a collection of
commands whose interfaces are statically tracked by the type
system. However, Frank eliminates the need for an additional effect
handling construct by generalising the basic mechanism of functional
abstraction itself. A function is simply the special case of a Frank
\emph{operator} that interprets no commands.
Moreover, Frank's operators can be \emph{multihandlers} which simultaneously
interpret commands from several sources at once, without disturbing
the direct style of functional programming with values.

Effect typing in Frank employs a novel form of effect polymorphism
which avoid mentioning effect variables in source code. This is
achieved by propagating an \emph{ambient ability} inwards, rather than
accumulating unions of potential effects outwards.

We introduce Frank by example, and then give a formal account of the
Frank type system and its semantics. We introduce Core Frank by
elaborating Frank operators into functions, case expressions, and
unary handlers, and then give a sound small-step operational semantics
for Core Frank.

Programming with effects and handlers is in its infancy. We contribute
an exploration of future possibilities, particularly in
combination with other forms of rich type system.
\end{abstract}



\section{Introduction}
\label{sec:intro}

\begin{center}
\begin{tabular}{c}
{\em Shall I be pure or impure?}$\qquad \qquad \qquad \qquad \qquad$\\
\hfill ---Philip Wadler~\cite{Wadler92b}
\end{tabular}
\end{center}

We say `Yes.': purity is a choice to make \emph{locally}. We introduce
\textbf{Frank}, an applicative language where the meaning of `impure'
computations is open to negotiation, based on Plotkin and Power's
algebraic effects~\cite{PlotkinP01a, PlotkinP01b, PlotkinP02,
  PlotkinP03} in conjunction with Plotkin and Pretnar's handlers for
algebraic effects~\cite{PlotkinP13}---a rich foundation
for effectful programming.
By separating effect interfaces from their implementation, algebraic
effects offer a high degree of modularity. Programmers can
express effectful programs independently of the concrete
interpretation of their effects. A handler gives one interpretation of the
effects of a computation.
In Frank, effect types (sometimes called simply \emph{effects} in the
literature) are known as \emph{abilities}. An ability denotes the
permission to invoke a particular set of commands.

Frank programs are written in direct style in the spirit of effect
type systems~\cite{LucassenG88, TalpinJ94}.
Frank \emph{operators} generalise call-by-value functions in two
dimensions.
First, operators handle effects. A unary operator is an effect
handler, acting as an interpreter for a specified set of commands
whose types are statically tracked by the type system. A unary
function is simply the special case of a unary operator whose handled
command set is empty.
Second, operators are $n$-ary, handling multiple computations over
distinct command sets simultaneously. An $n$-ary function is simply
the special case of an $n$-ary operator whose handled command sets are
all empty.

The contributions of this paper are:
\begin{itemize}
\item the definition of Frank, a strict functional programming
  language featuring a bidirectional effect type system, effect
  polymorphism, and effect handlers;
\item operators as both \emph{multihandlers} for handling multiple
  computations over distinct effect sets simultaneously and as
  \emph{functions} acting on values;
\item a novel approach to effect polymorphism which avoids mentioning
  effect variables in source code, crucially relying on the
  observation that one must always instantiate the effects of an
  operator being applied with the \emph{ambient ability}, that is,
  precisely those algebraic effects permitted by the current typing
  context;
\item a description of pattern matching compilation from Frank into a
  fairly standard call-by-value language with unary effect handlers,
  \emph{Core Frank};
\item a straightforward small-step operational semantics for Core
  Frank and a proof of type soundness;
\item an exploration of directions for future research, combining
  effect-and-handlers programming with features including
  substructural typing, dependent types, and totality.
\end{itemize}

A number of other languages and libraries are built around effect
handlers and algebraic effects.
Bauer and Pretnar's Eff~\cite{BauerP15} language is an ML-like
language extended with effect handlers. A significant difference
between Frank and the original version of Eff is that
the latter provides no support for effect typing. Recently Bauer and
Pretnar have designed an effect type system for
Eff~\cite{BauerP13}. Their implementation~\cite{Pretnar13} supports
Hindley-Milner type inference and the type system incorporates effect
subtyping.

Hillerstr\"om and
Lindley~\cite{Hillerstrom15,HillerstromL16,Hillerstrom16}
(Links~\cite{CooperLWY06}) and Leijen~\cite{Leijen17}
(Koka~\cite{Leijen14}) have extended existing languages with effect
handlers and algebraic effects. Both languages incorporate row-based
effect type systems and attempt to elide some effect variables from
source code, but neither eliminates effect variables to the extent
that Frank does.
Dolan et al.~\cite{Dolan2015} built Multicore OCaml by extending OCaml
with support for effect handlers and algebraic effects. Multicore
OCaml does not include an effect type system.

Whereas Frank is bidirectionally typed, all of these other languages
use Hindley-Milner type inference.
None of the other languages supports multihandlers and none of those
with effect typing allow effect variables to be omitted to the degree
that Frank does.

Kammar et al.~\cite{KammarLO13} describe a number of effect handler
libraries for languages ranging from Racket, to SML, to OCaml, to
Haskell. Apart from the Haskell library, their libraries have no
effect typing support. The Haskell library takes advantage of type
classes to simulate an effect type system not entirely dissimilar to
that of Frank. As Haskell is lazy, the Haskell library cannot be used
to write direct-style effectful programs---one must instead adopt a
monadic style. Moreover, although there are a number of ways of almost
simulating effect type systems in Haskell, none is without its
flaws. Kiselyov and collaborators~\cite{KiselyovSS13,KiselyovI15} have
built another Haskell library for effect handlers, making different
design choices.

Brady's \verb!effects! library~\cite{Brady13} provides a DSL for
programming with effects in the dependently typed language
Idris~\cite{BradyIdris13}. Like the Haskell libraries, Brady's library
currently requires the programmer to write effectful code in a monadic
style.

McBride's Shonky~\cite{McBride16} is essentially an untyped version of
Frank, with a somewhat different concrete syntax.
We have built a prototype implementation of Frank by translating typed
Frank programs into Shonky. The implementation is available at the
following URL:

\begin{center}
\url{https://www.github.com/frank-lang/frank}
\end{center}

%
%

The rest of the paper is structured as follows.
Section~\ref{sec:examples} introduces Frank by
example. Section~\ref{sec:frank} presents abstract syntax and a type
system for Frank. Section~\ref{sec:core} describes how to elaborate
operators into Core Frank, a language of plain call-by-value
functions, explicit case analysis, and unary handler constructs.
%
Section~\ref{sec:semantics} gives an operational semantics for Core Frank and
proves type soundness. Section~\ref{sec:compdata} discusses how to store
computations in data structures. Section~\ref{sec:implementation} describes
the status of our implementation. Section~\ref{sec:related} outlines related
work. Section~\ref{sec:future} discusses future work and Section~\ref{sec:con}
concludes.

\section{A Frank Tutorial}
\label{sec:examples}

\begin{center}
\begin{tabular}{l}
{\em `To be is to do'---Socrates.}\\
{\em `To do is to be'---Sartre.}\\
{\em `Do be do be do'---Sinatra.}$\qquad\qquad\qquad\qquad\qquad$\\
\hfill---anonymous graffiti, via Kurt Vonnegut~\cite{Vonnegut82}
\end{tabular}
\end{center}

Frank is a functional programming language with effects and handlers
in the spirit of Eff~\cite{BauerP15}, controlled by a type system
inspired by Levy's call-by-push-value~\cite{Levy2004}.
Doing and Being are clearly separated, and managed by distinguished
notions of computation and value types.

\subsection{Data Types and First-Order Functions}

Concrete values live in inductive data types. By convention (not
compulsion), we give type constructors uppercase initials, and may
apply prefixed to parameters, also written uppercase. Data
constructors are prefix and, again by convention, initially lowercase.
%
\begin{verbatim}
data Zero =
data Unit = unit
data Bool = tt | ff

data Nat    = zero | suc Nat
data List X = nil  | cons X (List X)

data Pair X Y = pair X Y
\end{verbatim}
We choose to treat constructors as distinct from functions, and
constructors must always be fully applied.

We can write perfectly ordinary first-order functional programs by
pattern matching. Type signatures are compulsory, universally
quantifying implicitly over freely occurring type variables, and
insisting on the corresponding parametric polymorphism.
\begin{verbatim}
append : List X -> List X -> List X
append nil         ys = ys
append (cons x xs) ys = cons x (append xs ys)
\end{verbatim}

\subsection{Effect Polymorphism in Ambient Silence}

Computations, such as functions, have computation types, which embed
explicitly into the value types: braces play the role of `suspenders'
in types and values. Accordingly, we can write typical higher-order
functions
\begin{verbatim}
map : {X -> Y} -> List X -> List Y
map f nil         = nil
map f (cons x xs) = cons (f x) (map f xs)
\end{verbatim}
and apply them in the usual way:
\begin{verbatim}
map {n -> n+1} (cons 1 (cons 2 (cons 3 nil)))
  = cons 2 (cons 3 (cons 4 nil))
\end{verbatim}

A value type \verb!A! is a data type \verb!D R1 ... Rn!, a suspended
computation type \verb!{C}!, or a type variable \verb!X!.
A computation type resembles a function type
$\verb!T1 -> ... -> Tm -> [I1 ... In]B!$ with $m$ \emph{ports} and a
\emph{peg} showing the \emph{ability} the computation needs---a
bracketed list of $n$ enabled \emph{interfaces}---and the \emph{value
  type} it delivers.
In Frank, names always bind values (a simplifying decision which we
shall re-examine in section \ref{sec:future}). Top level definitions
give names to suspended computations, but we omit the outer braces in
their types for convenience.

Type checking separates cleanly into checking the compatibility
of value types and checking that required abilities are available.
Empty brackets may be omitted. We could have written
\begin{verbatim}
map : {X -> []Y} -> List X -> []List Y
\end{verbatim}
which really means
\begin{verbatim}
map : {{X -> []Y} -> List X -> []List Y}
\end{verbatim}
but have a care: the empty bracket stands for the \emph{ambient}
ability, not for purity; the \verb!map! operator is implicitly
effect-polymorphic.

The type of \verb!map! in Frank says that whatever ability an instance
receives will be offered in turn to the operator that acts on each
element. That is, we have written something like ML's \verb|map| but
without giving up control over effects, and we have written something
like Haskell's \verb!map! but acquired a function as general as its
monadic \verb!mapM!, as we shall see just as soon as we acquire
nontrivial ability.

\subsection{Controlling Evaluation}

Frank is a (left-to-right) call-by-value language, so we should be
careful when defining control operators. For instance, we may define
sequential composition operators
\begin{verbatim}
fst : X -> Y -> X          snd : X -> Y -> Y
fst x y = x                snd x y = y
\end{verbatim}
Both arguments are evaluated (relative to the ambient ability), before
one value is returned. We take the liberty of writing \verb!snd x y!
as \verb!x; y!, echoing the ML semicolon, and note its associativity.

Meanwhile, avoiding evaluation must be done explicitly by suspending
computations. The following operator
\begin{verbatim}
iffy : Bool -> X -> X -> X
iffy tt t f = t
iffy ff t f = f
\end{verbatim}
is the conditional expression operator which forces evaluation of the
condition and \emph{both} branches, before choosing between the
values. To write the traditional conditional, we must therefore
suspend the second and third arguments:
\begin{verbatim}
if : Bool -> {X} -> {X} -> X
if tt t f = t!
if ff t f = f!
\end{verbatim}
Again, Frank variables stand for values, but \verb!t! and \verb!f!
  are not values of type \verb!X!. Rather, they \emph{are} suspended
  computations of type \verb!{X}!, but we must \emph{do} just one.
The postfix \verb$!$ denotes nullary application of a suspended
computation.
%


We write suspended computations in braces, with a choice of zero or
more pattern matching clauses separated by \verb!|! symbols. In a
nullary suspension, we have one choice, which is just written as an
expression in braces, for instance,
\begin{verbatim}
if fire! {launch missiles} {unit}
\end{verbatim}
assuming that \verb!launch! is a command permitted by the ambient
ability, granted to both branches by the silently effect-polymorphic
type of \verb!if!.

With non-nullary suspensions we can simulate case-expressions inline
using reverse application
\begin{verbatim}
on : X -> {X -> Y} -> Y
on x f = f x
\end{verbatim}
as in this example of the short-circuited `and':
\begin{verbatim}
shortAnd : Bool -> {Bool} -> Bool
shortAnd x c = on x { tt -> c! | ff -> ff }
\end{verbatim}

\subsection{Abilities Collect Interfaces; Interfaces Offer Commands}

Abilities (Frank's realisation of algebraic effects) are collections
of parameterised interfaces, each of which describes a choice of
commands (known elsewhere as
\emph{operations}~\cite{PlotkinP13}). Command types may refer to the
parameters of their interface but are not otherwise polymorphic.
Here are some simple interfaces.
\begin{verbatim}
interface Send X     = send : X -> Unit

interface Receive X  = receive : X

interface State S    = get : S
                     | put : S -> Unit

interface Abort      = aborting : Zero
\end{verbatim}
The \verb!send! command takes an argument of type \verb!X! and returns
a value of type \verb!Unit!. The \verb!receive! command returns a
value of type \verb!X!. The \verb!State! interface offers \verb|get|
and \verb|set| commands.  Note that, unlike data constructors,
commands are first-class values. In particular, while \verb!Zero! is
uninhabited, \verb!{[Abort]Zero}! contains the value
\verb!aborting!. Correspondingly, we can define a polymorphic
\verb!abort! which we can use whenever \verb!Abort! is enabled
\begin{verbatim}
abort : [Abort]X
abort! = on aborting! {}
\end{verbatim} 
by empty case analysis.
The postfix \verb$!$ attached to \verb$abort$ denotes the definition
of a nullary operator.

We may use the silent effect polymorphism of \verb!map! nontrivially
to send a list of elements, one at a time:
\begin{verbatim}
sends : List X -> [Send X]Unit
sends xs = map send xs; unit
\end{verbatim}
The reason this type checks at all is because \verb!map! is implicitly
polymorphic in its effects. The bracket \verb![Send X]! demands that
the ambient ability permits \emph{at least} the \verb!Send X!
commands. The type of \verb!map! works with \emph{any} ambient
ability, hence certainly those which permit \verb!Send X!, and it
passes that ability to its computation argument, which may thus be
\verb!send!.

However, the following does not typecheck, because \verb!Send X! has
not been included in the peg of \verb!bad!.
\begin{verbatim}
bad : List X -> Unit
bad xs = map send xs; unit
\end{verbatim}

There is no effect inference in Frank. The typing rules' conclusions
do not accumulate the abilities of the programs in their
premisses. Rather, we are explicit about what the environment makes
possible---the ambient ability---and where and how that changes.

In designing Frank we have sought to maintain the benefits of effect
polymorphism whilst avoiding the need to write effect variables in
source code.
There are no explicit effect variables in any of the examples in this
paper.
In an earlier draft of this paper we ruled out explicit effect
variables by fiat.
But doing so placed artificial restrictions on the formalism (see
Section~\ref{sec:frank}), so we do now permit them. A case where
explicit effect variables may be useful is in manipulating data types
containing multiple suspended computations with different abilities;
we are yet to explore compelling use cases.

\subsection{Direct Style for Monadic Programming}

We work in a direct applicative style. Where the output of one
computation is used as the input to another, we may just write an
application, or a case analysis, directly. For instance, we can
implement the result of repeatedly reading lists until one is empty
and concatenating the result.
\begin{verbatim}
catter : [Receive (List X)]List X
catter! = on receive! { nil -> nil
                      | xs  -> append xs catter! }
\end{verbatim}

In Haskell, \verb|receive!| would be a monadic computation \verb|ask|
unsuitable for case analysis---its value would be extracted and named
before inspection, thus:
\begin{verbatim}
catter :: Reader (List a) (List a)   -- Haskell
catter = do               
  xs <- ask
  case xs of
    [] -> return []
    xs -> do ys <- catter; return (xs ++ ys)
\end{verbatim}
The latter part of \verb!catter! could perhaps be written without
naming \verb!ys! as \verb!(xs ++) <$> catter!, or even, with `idiom
brackets', \verb!(|pure xs ++ catter|)!, but always there is extra
plumbing (here \verb!do!-notation and \verb!return!) whose only
purpose is to tell the compiler where to parse a type as
$\mathit{effect}\:\mathit{value}$ and where just as $\mathit{value}$.
The choice to be frank about the separation of effects from values in
the syntax of types provides a stronger cue to the status of each
component and reduces the need for plumbing.
We do not, however, escape the need to disambiguate \emph{doing}
\verb|receive!| from \emph{being} \verb|receive|.

%

In the same mode, we can implement the \verb!C++! `increment \verb!c!,
return original value' operation as follows.
\begin{verbatim}
next : [State Int]Int
next! = fst get! (put (get! + 1))
\end{verbatim}
In Haskell \verb!next! would have to be explicitly sequentialised.
\begin{samepage}
\begin{verbatim}
next :: State Int Int
next = do x <- get
          y <- get
          put y+1
          return x
\end{verbatim}
\end{samepage}
(We have written \verb!get! twice to match the preceding Frank code,
but assuming a standard implementation of state one could of course
delete the second \verb!get! and use \verb!x! in place of \verb!y!.)
The absence of explicit plumbing in Frank depends crucially on the
fact that Frank, unlike Haskell, has a fixed evaluation order.

\subsection{Handling by Application}

In a call-by-value language a function application can be seen as a
particularly degenerate mode of coroutining between the function and
its argument. The function process waits while the argument process
computes to a value, transmitted once as the argument's terminal
action; on receipt, the function post-processes that value in some
way, before transmitting its value in turn.



Frank is already distinct from other languages with effect handlers in
its effect type system, but the key departure it makes in program
style is to handle effects without any special syntax for invoking an
effect handler. Rather, the ordinary notion of `function' is extended
with the means to offer effects to arguments, invoked just by
application. That is, the blank space application notation is used for
more general modes of coroutining between operator and arguments than
the return-value-then-quit default. For instance, the usual behaviour
of the `state' commands can be given as follows.
\begin{verbatim}
state : S -> <State S>X -> X
state _ x            = x
state s <get -> k>   = state s (k s)
state _ <put s -> k> = state s (k unit)
\end{verbatim}
Let us give an example using \verb!state! before unpacking its
definition. We might pair the elements of a list with successive
numbers.
\begin{verbatim}
index : List X -> List (Pair Int X)
index xs = state 0 (map {x -> pair next! x} xs)
\end{verbatim}

Allowing string notation for lists of characters we obtain:
\begin{verbatim}
index "abc" = cons (pair 0 'a')
               (cons (pair 1 'b')
                 (cons (pair 2 'c') nil))
\end{verbatim}
What is happening?

The type of \verb!state! shows us that Frank operators do not merely
have input \emph{types}, but input \emph{ports}, specifying not only
the types of the values expected, but also an \emph{adjustment} to the
ambient ability, written in chevrons and usually omitted when it is
the identity (as in all of the examples we have seen so far). Whatever
the ambient ability might be when \verb!state! is invoked, the initial
state should arrive at its first port using only that ability; the
ambient ability at its second port will include the \verb!State S!
interface, shadowing any other \verb!State A! interfaces which might
have been present already. Correspondingly, by the time \verb!index!
invokes \verb!map!, the ambient ability includes \verb!State Int!,
allowing the elementwise operation to invoke \verb|next!|.

%
The first equation of \verb!state! explains what to do if any
\emph{value} arrives on the second port. In Frank, a traditional
pattern built from constructors and variables matches only
\emph{values}, so the \verb!x! is not a catch-all pattern, as things
other than values can arrive at that port. In particular,
\emph{requests} can arrive at the second port, in accordance with the
\verb!State S! interface. A request consists of a command instance and
a continuation. Requests are matched by patterns in chevrons which
show the particular command instance being handled left of \verb!->!,
with a pattern variable standing for the \emph{continuation} on the
right. The patterns of \verb!state! thus cover all possible \emph{signals}
(that is, values or requests) advertised as acceptable at its ports.

Having received signals for each argument the \verb!state! operator
should \emph{handle} them.
If the input on the second port is a value, then that value is
returned.
If the input on the second port is a request, then the \verb!state!
  operator is reinvoked with a new state (which is simply the old
  state in the case of \verb!get! and \verb!s! in the case of
  \verb!put s!) in the first port and the continuation \emph{invoked}
  in the second port.

We emphasise a key difference between Frank and most other languages
supporting algebraic effect handlers (including Eff, Koka, and
Multicore OCaml~\cite{Dolan2015}): Frank's continuation variables are
shallow in that they capture only the rest of the subordinated
computation, not the result of handling it, allowing us to change how
we carry on handling, for instance, by updating the state.
In contrast, Multicore OCaml's continuation variables are deep in that
invoking them implicitly reinvokes the handler.
Consider the definition for \verb!state! in Multicore OCaml
\begin{verbatim}
effect Put : t -> unit
let put x = perform (Put x)

effect Get : t
let get () = perform Get

let state m =
  match m () with
  | x                -> fun s -> x
  | effect Get     k -> fun s -> continue k s s
  | effect (Put s) k -> fun _ -> continue k () s
\end{verbatim}
Multicore OCaml provides three special keywords for algebraic effects
and handlers: \verb!effect! declares a command or marks a request
pattern, \verb!perform! invokes a command, and \verb!continue!
invokes a continuation.
The shallow implementation of \verb!state! in Frank requires explicit
recursion. The deep implementation of \verb!state! in Multicore OCaml
performs the recursion implicitly. On the other hand, the shallow
version allows us to thread the state through the operator, whereas
the deep version relies on interpreting a stateful computation as a
function and threading the state through the continuation.


Shallow handlers can straightforwardly express deep handlers using
explicit recursion.
Deep handlers can encode shallow handlers in much the same way that
iteration (catamorphism, fold) can encode primitive recursion
(paramorphism), and with much the same increase in complexity. On the
other hand, handlers which admit a deep implementation have a more
regular behaviour and admit easier reasoning, just as `folds' offer
specific proof techniques not available to pattern matching programs
in general.
\citeauthor{KammarLO13}~\cite{KammarLO13} provide a more in-depth
discussion of the trade-offs between deep and shallow handlers.

\subsection{Handling on Multiple Ports}
\label{subsec:pipe}

Frank allows the programmer to write $n$-ary operators, so we can
offer different adjustments to the ambient ability at different
ports. For instance, we can implement a \verb!pipe! operator which
matches \verb!receive! commands downstream with \verb!send! commands
upstream.
\begin{verbatim}
pipe : <Send X>Unit -> <Receive X>Y -> [Abort]Y
pipe <send x -> s> <receive -> r> =
  pipe    (s unit) (r x)  
pipe <_>           y              = y
pipe unit          <_>            = abort!
\end{verbatim}
The type signature conveys several different things. The \verb!pipe!
  operator must handle all commands from \verb!Send X! on its first
  port and all commands from \verb!Receive X! on its second port. We
  say that \verb!pipe! is thus a \emph{multihandler}. The first
  argument has type \verb!Unit! and the second argument has type
  \verb!Y!. The operator itself is allowed to perform \verb!Abort!
  commands and returns a final value of type \verb!Y!.

The first line implements the communication between producer and
consumer, reinvoking \verb!pipe! with both continuations, giving the
sent value to the receiver. The second line makes use of the catch-all
pattern \verb!<_>! which matches \emph{either} a \verb!send! command
or an attempt to return a value: as the consumer has delivered a value
the producer can be safely discarded. The third line covers the case
which falls through: the catch-all pattern must be a \verb!receive!
  command, as the value case has been treated already, but the
  producer has stopped sending, so \verb!abort! is invoked to indicate
  a `broken pipe'.

We can run \verb!pipe! as follows:
\begin{verbatim}
pipe (sends (cons "do" (cons "be" (cons "" nil))))
     catter!
  = "dobe"
\end{verbatim}
Moreover, if we write
\begin{verbatim}
spacer : [Send (List Char),
          Receive (List Char)]Unit
spacer! = send receive; send " "; spacer!
\end{verbatim}
we find instead that
\begin{verbatim}
pipe (sends (cons "do" (cons "be" (cons "" nil))))
     (pipe spacer! catter!)
  = "do be "
\end{verbatim}
where the \verb!spacer!'s \verb!receive!s are handled by the outer
\verb!pipe!, but its \verb!send!s are handled by the inner one. The
other way around also works as it should, that is, \verb!pipe! is
associative.
\begin{verbatim}
pipe (pipe
      (sends (cons "do" (cons "be" (cons "" nil))))
      spacer!) catter!
  = "do be "
\end{verbatim}

There is nothing you can do with simultaneous handling that you cannot
also do with mutually recursive handlers for one process at a
time. The Frank approach is, however, more direct.
\citeauthor{KammarLO13}~\cite{KammarLO13} provide both deep and
shallow handler implementations for pipes using their Haskell effects
library. Both implementations are significantly more complex than the
above definition in Frank, requiring the unary handler for sending
(receiving) to maintain a suspended computation to the consumer
(producer) to continue the interaction upon receipt of a
command. Moreover, the deep handler implementation depends on a
non-trivial mutually recursive data type, which places considerable
cognitive load on the programmer. So, even in systems such as
Multicore OCaml and Eff, offering a more aesthetic syntax than
\citeauthor{KammarLO13}'s library, a programming burden remains.

Let us clarify that the adjustment marked in chevrons on a port
promises \emph{exactly} what will be handled at that port.
The peg of \verb!pipe! requires the ambient ability to support
\verb!Abort!, and its ports offer to extend that ability with
\verb!Send X! and \verb!Receive X!, respectively, so the producer and
consumer will each also support \verb!Abort!. However, because neither
port advertises \verb!Abort! in its adjustment, the implementation of
\verb!pipe! may not intercept the \verb!aborting! command. In
particular, the catch-all pattern \verb!<_>! matches only the signals
advertised at the relevant port, with other commands forwarded
transparently to the most local port offering the relevant
interface. No Frank process may secretly intercept commands. Of
course, the \verb!pipe! operator can prevent action by ignoring the
continuation to a \verb!send! on its first port or a \verb!receive! on
its second, but it cannot change the meaning of other commands.

One can simulate adjustments using a more conventional effect type
system with abilities on both ports and pegs. However, this yields
more verbose and less precise types.
For instance, the type of the first argument to \verb!pipe! becomes
\verb!<Abort, Send X>Unit! instead of \verb!<Send X>Unit!. The type of
the port has been polluted by the ability of the peg and it now fails
to advertise precisely which interfaces it handles.

\subsection{The Catch Question}


Frank allows us to implement an `exception handler' with a slightly
more nuanced type than is sometimes seen.
\begin{verbatim}
catch : <Abort>X -> {X} -> X
catch x               _ = x
catch <aborting -> _> h = h!
\end{verbatim}
The first argument to \verb!catch! is the computation to run that may
raise an exception. The second argument is the alternative computation
to run in the case of failure, given as a suspended computation
allowing us to choose whether to run it.
We do not presume that the ambient ability in which \verb!catch! is
executed offers the \verb!Abort! interface.
In contrast, a typical treatment of exceptions renders \verb!catch! as
the prioritised choice between two failure-prone computations.
For instance, the Haskell mtl library offers
\begin{verbatim}
catchError :: -- Haskell
   MonadError () m => m a -> (() -> m a) -> m a
\end{verbatim}
where the exception handler is always allowed to throw an error. In
other words, this Haskell typing unnecessarily makes the ability to
abort non-local.
Leijen makes a similar observation in Koka's treatment of
exceptions~\cite{Leijen14}.

Frank's effect polymorphism ensures that the alternative
computation is permitted to abort if and only if \verb!catch!
is, so we lose no functionality but gain precision. Moreover, we
promise that \verb!catch! will trap \verb!aborting!
only in its first port, so that any failure (or anything else) that
\verb|h!| does is handled by the environment---indeed, you can see that
\verb|h!| is executed as a tail call, if at all, thus outside the
scope of \verb!catch!.
In the case that the ambient is allowed to abort, then when the
adjustment is applied to the ambient ability we obtain an ability with
two instances of the \verb!Abort! interface. Just like Koka, Frank
resolves any replication of interfaces by shadowing, discarding all
but the last instance of an interface in an ability.

\subsection{The Disappearance of Control}

Using one of the many variations on the theme of free monads, we could
implement operators like \verb!state!, \verb!pipe! and \verb!catch!
  as abstractions over \emph{computations} reified as command-response
  trees. By contrast, our handlers do not abstract over computations,
  nor do they have computation-to-computation handler types distinct
  from value-to-computation function types~\cite{BauerP13,
    KammarLO13}.

Frank computations are abstract: a thing of type \verb!{C}! can be
communicated or invoked, but not inspected. Ports explain which values
are expected, and operators match on those values directly, without
apparently forcing a computation, yet they also admit other
specific modes of interaction, handled in specific ways.

Semantically, then, a Frank operator must map computation trees to
computation trees, but we write its action on values directly and its
handling of commands minimally. The machinery by which commands from
the input not handled locally must be forwarded with suitably wrapped
continuations is hard-wired, as we shall make explicit in
Sections~\ref{sec:core}~and~\ref{sec:semantics}.

However, let us first give the type system for these programs and show
how Frank's careful silences deliver the power we claim.

\section{A Frank Formalism}
\label{sec:frank}

\begin{figure}
\fighead{Types}
\begin{syntax}
\slab{value types}       &A, B          &::=& D~\many{R} \mid \thunk{C} \mid X \\
\slab{computation types} &C             &::=& \many{T \to}~G \\
\slab{ports}        &T             &::=& \effin{\adj}A \\
\slab{pegs}         &G             &::=& \effout{\sigs}A \\
[1ex]
\slab{type variables}& Z          &::=& X \mid [\ev] \\
\slab{type arguments}& R           &::=& A \mid [\sigs] \\
\slab{polytypes}    &P             &::=& \forall \many{Z}.A \\
[1ex]
\slab{abilities}    &\sigs  &::=&
  \nowt \mid \sigs, \sig~\many{R} \mid \ev \\
\slab{adjustments}  &\adj  &::=&
  \id \mid \adj + \sig~\many{R} \\
[1ex]
\slab{type environments}  &\Gamma        &::=& \cdot \mid \Gamma, x:A \mid f:P \\
\end{syntax}

\fighead{Terms}
\begin{syntax}
\slab{uses}               &m       &::=& x \mid f \mid c \mid m~s \\
\slab{constructions}      &n       &::=& m \mid k~\many{n} \mid \thunk{e} \\
                       &       &\mid& \key{let}~f : P = n~\key{in}~n' \\
                       &       &\mid& \key{letrec}~\many{f : P = e}~\key{in}~n \\
[1ex]
\slab{spines}          &s       &::=& \many{n} \\
\slab{computations}     &e       &::=& \many{\many{r} \mapsto n}
\\[1ex]
\slab{computation patterns}&r &::=& p
                               \mid \effin{\handle{c~\many{p}\,}{z}}
                               \mid \effin{x} \\
\slab{value patterns}&p       &::=& k~\many{p} \mid x        \\
\end{syntax}

\caption{Frank Abstract Syntax}
\label{fig:frank-syntax}
\end{figure}

\begin{center}
\begin{tabular}{c}
{\em A value is. A computation does.} $\qquad \qquad \qquad $\\
\hfill---Paul Blain Levy~\cite{Levy2004}
\end{tabular}
\end{center}
In this section we give a formal presentation of the abstract syntax
and type system of Frank.

\subsection{Syntax}
The abstract syntax of Frank is given in
Figure~\ref{fig:frank-syntax}.


The types are divided into value types and computation types.
%
Value types are data types ($D~\many{R}$),
%
%
suspended computation types
($\thunk{C}$), or type variables ($X$).

Computations types are build from input \emph{ports} $T$ and output
\emph{pegs} $G$.
A computation type
\[
C = \effin{\adj_1}A_1 \to \dots \to \effin{\adj_n}A_n \to \effout{\sigs}B
\]
has ports $\effin{\adj_1}A_1, \dots, \effin{\adj_n}A_n$ and
peg $\effout{\sigs}B$. 
A computation of type $C$ must handle effects in $\adj_i$ on the
$i$-th argument. All arguments are handled simultaneously. As a result
it returns a value of type $B$ and may perform effects in $\sigs$.

A port $\effin{\adj}A$ constrains an input. The adjustment $\adj$
describes the difference between the ambient effects and the effects
of the input, in other words, those effects occurring in the input
that must be handled on that port.
A peg $\effout{\sigs}A$ constrains an output. The effects $\sigs$ are
those that result from running the computation.

\paragraph{Effect Polymorphism with an Invisible Effect Variable}

Consider the type of \verb!map! in Section~\ref{sec:examples}:
\[
\thunk{X \to Y} \to \var{List}~X \to \var{List}~Y
\]
Modulo the braces around the function type, this is the same type a
functional programmer might expect to write in a language without
support for effect typing.
In fact, this type desugars into:
\[
\effin{\id}\thunk{\effin{\id}X \to \effout{\evd}Y}
  \to \effin{\id}(\var{List}~X) \to \effout{\evd}(\var{List}~Y)
\]
We adopt the convention that the identity adjustment $\id$ may be
omitted from adjustments and ports.
\begin{equations}
\sig_1~\many{R_1}, \dots, \sig_n~\many{R_n} &\equiv& \id + \sig_1~\many{R_1} + \dots + \sig_n~\many{R_n} \\
A &\equiv& \effin{\id}A \\
\end{equations}
Similarly, we adopt the convention that effect variables may be
omitted from abilities and pegs.
\begin{equations}
\sig_1~\many{R_1}, \dots, \sig_n~\many{R_n} &\equiv& \evd, \sig_1~\many{R_1}, \dots, \sig_n~\many{R_n} \\
A &\equiv& \effout{\evd}A \\
\end{equations}
Here $\evd$ is a distinguished effect variable, the \emph{implicit
  effect variable} that is fresh for every type signature in a
program. This syntactic sugar ensures that we need never write the
implicit effect variable $\evd$ anywhere in a Frank program.


\medskip




%
We let $X$ range over ordinary type variables and $\ev$ range over
effect variables; polytypes may be polymorphic in both. Though we
avoid effect variables in source code, we are entirely explicit about
them in the abstract syntax and the type system.


\newcommand{\dcon}{\mathcal{D}}
\newcommand{\icom}{\mathcal{I}}

Data Types and effect interfaces are defined globally.
A definition for data type $D(\many{Z})$ consists of a collection of
data constructor signatures of the form $k:\many{A}$, where the
type/effect variables $\many{Z}$ may be bound in the data constructor
arguments $\many{A}$. Each data constructor belongs to a single data
type and may appear only once in that data type.
We write $\dcon(D~\many{R}, k)$ for the type arguments of constructor
$k$ of data type $D~\many{R}$.
%
A definition for effect interface $I(\many{Z})$ consists of a
collection of command signatures of the form $c:\many{A} \to B$,
denoting that command $c$ takes arguments of types $\many{A}$ and
returns a value of type $B$. The types $\many{A}$ and $B$ may all
depend on $\many{Z}$. Each command belongs to a single interface and
may appear only once in that interface.
We write $\icom(I~\many{R}, c)$ for the signature of command $c$ of
effect interface $I~\many{R}$

\paragraph{Effect Parameters with an Invisible Effect Variable}
In the case that the first parameter of a data type or effect
interface definition is its only effect variable $\evd$, then we may
omit it from the definition (we give an example in
Section~\ref{sec:compdata}).
\medskip

An ability is a collection of interfaces initiated either with the
empty ability $\emptyset$ (yielding a \emph{closed} ability) or an
effect variable $\ev$ (yielding an \emph{open} ability). Order is
important, as repeats are permitted, in which case the right-most
interface overrides all others with the same name.
Closed abilities are not normally required, but they can be used to
enforce purity, for instance. In ASCII source code we write
$\emptyset$ as \verb!0!.

Adjustments modify abilities. The identity adjustment $\id$ leaves an
ability unchanged. An adjustment $\adj + \sig~\many{R}$ extends an
ability with the interface $\sig~\many{R}$. The action of an
adjustment $\Delta$ on an ability $\sigs$ is given by the $\oplus$
operation.
\begin{equations}
\sigs \oplus \id                    &=& \sigs \\
\sigs \oplus (\adj + \sig~\many{R}) &=& (\sigs \oplus \adj), \sig~\many{R} \\
\end{equations}

Type environments distinguish monomorphic and polymorphic variables.

Frank follows a bidirectional typing discipline~\cite{PierceT00}. Thus
terms are subdivided into \emph{uses} whose type may be inferred, and
\emph{constructions} which may be checked against a type. Uses
comprise monomorphic variables ($x$), polymorphic variables ($f$),
commands ($c$), and applications ($m~s$). Constructions comprise uses
($m$), data constructor instances ($k~\many{n}$), suspended
computations ($\thunk{e}$), polymorphic let ($\key{let}~f : P =
n~\key{in}~n'$) and mutual recursion ($\key{letrec}~\many{f : P =
  e}~\key{in}~n$).
A spine ($s$) is a sequence of constructions ($\many{n}$). We write
$!$ for the empty spine.

A computation is defined by a sequence of pattern matching clauses
($\many{\many{r} \mapsto n}$).
Each pattern matching clause takes a sequence of computation patterns
($\many{r}$). A computation pattern is either a standard value pattern
($p$), a request pattern ($\effin{\handle{c~\many{p}\,}{z}}$), which
matches command $c$ binding its arguments to $\many{p}$ and the
continuation to $z$, or a catch-all pattern $\effin{x}$, which matches
any value or handled command, binding it to $x$.
A value pattern is either a data constructor pattern ($k~\many{p}$) or
a variable pattern $x$.

\begin{figure*}
$\boxed{\infers{\Gamma}{\sigs}{m}{A}}$
\begin{mathpar}
\inferrule[Var]
  {x:A \in \Gamma}
  {\infersgs{x}{A}}

\inferrule[PolyVar]
  {f:\forall \many{Z}.A \in \Gamma}
  {\infersgs{f}{\theta(A)}}

\inferrule[Command]
  {c : \many{A \to}~ B \in \sigs}
  {\infersgs{c}{\thunk{\many{\effin{\id}A \to}~ \effout{\sigs}B}}}

\inferrule[App]
  {\infersgs{m}{\thunk{\many{\effin{\adj}A \to}~ \effout{\sigs'}B}}\\
   \sigs' = \sigs \\
   \many{\checks{\Gamma}{\sigs \oplus \adj}{A}{n}}}
  {\infers{\Gamma}{\sigs}{m~\many{n}}{B}}
\end{mathpar}

$\boxed{\checks{\Gamma}{\sigs}{A}{n}}$

\begin{mathpar}
\inferrule[Switch]
  {\infersgs{m}{A} \\ A = B}
  {\checksgs{B}{m}}

\inferrule[Data]
  {k~\many{A} \in D~\many{R} \\
   \many{\checksgs{A}{n}}}
  {\checksgs{D~\many{R}}{k~\many{n}}}

\inferrule[Thunk]
  {\checksdefg{C}{e}}
  {\checksgs{\thunk{C}}{\thunk{e}}}
\\
\inferrule[Let]
  {P = \forall \many{Z}.A \\
   \checksgs{A}{n} \\
   \checks{\Gamma, f : P}{\sigs}{B}{n'}}
  {\checksgs{B}{\key{let}~f : P = n~\key{in}~n'}}

\inferrule[LetRec]
  {\many{P} = \many{\forall \many{Z}.\thunk{C}} \\
   \many{\checksdef{\Gamma, \many{f : P}}{C}{e}} \\
   \checks{\Gamma, \many{f : P}}{\sigs}{B}{n}}
  {\checksgs{B}{\key{letrec}~\many{f : P = e}~\key{in}~n}}
\end{mathpar}

\begin{minipage}{.5\linewidth}
$\boxed{\checksdefg{C}{e}}$
\begin{mathpar}
\inferrule[Comp]
  {(\matchesc{T_j}{r_{i,j}}{\sigs}{\Gamma'_{i,j}})_{i,j} \\
   (\checks{\Gamma, (\Gamma'_{i,j})_j}{\sigs}{B}{n_i})_i \\\\
   (r_{i,j})_{i,j} \text{ covers } (T_j)_j}
  {\checksdefg{(T_j \to)_j~\effout{\sigs}B}{((r_{i,j})_j \mapsto n_i)_i}}
\end{mathpar}
\end{minipage}
\begin{minipage}{.5\linewidth}
$\boxed{\matchesv{A}{p}{\Gamma}}$

\begin{mathpar}
\inferrule[P-Var]
  { }
  {\matchesv{A}{x}{x:A}}

\inferrule[P-Data]
  {k~\many{A} \in D~\many{R} \\
   \many{\matchesv{A}{p}{\Gamma}}}
  {\matchesv{D~\many{R}}{k~\many{p}}{\many{\Gamma}}}
\end{mathpar}
\end{minipage}


\medskip

$\boxed{\matchesc{T}{r}{\sigs}{\Gamma}}$

\begin{mathpar}
\inferrule[P-Value]
  {\matchesv{A}{p}{\Gamma}}
  {\matchesc{\effin{\adj}A}{p}{\sigs}{\Gamma}}

\inferrule[P-Request]
  {c:\many{A} \to B \in \nowt \oplus \adj \\
   (\matchesv{A_i}{p_i}{\Gamma_i})_i}
  {\matchesc{\effin{\adj}B'}
           {\effin{\handle{c~\many{p}}{z}}}
           {\sigs}
           {\many{\Gamma}, z:{\effin{\id}B \to \effout{\sigs \oplus \adj}B'}}}

\inferrule[P-CatchAll]
  { }
  {\matchesc{\effin{\adj}A}{\effin{x}}{\sigs}{x:{\thunk{\effout{\sigs \oplus \adj}A}}}}
\end{mathpar}

\caption{Frank Typing Rules}
\label{fig:frank-typing}
\end{figure*}

\paragraph{Example}
To illustrate how source programs may be straightforwardly represented
as abstract syntax, we give the abstract syntax for an example
involving the \verb!map!, \verb!state!, and \verb!index!  operators
from Section~\ref{sec:examples}.
\[
\def\arraystretch{1.1}
\bl
\key{letrec}~\op{map} : \\
\quad
   \forall \evd~X~Y.
      \thunk{\effin{\id}\thunk{\effin{\id}X \to \effout{\evd}Y}
               \to \effin{\id}(\con{List}~X) \to \effout{\evd}(\con{List}~Y)} \\
\qquad =
       \ba[t]{@{}r@{}l@{}l@{}}
                           & f~\con{nil}               &~\mapsto \con{nil} \\
                           & f~(\con{cons}~x~\con{xs}) &~\mapsto \con{cons}~(f~x)~(\op{map}~f~\con{xs}) ~\key{in} \\
       \ea \\
\key{letrec}~\op{state} : \forall \evd~X.\thunk{\effin{\id}X \to \effin{\id + \inter{State}~S}X \to \effout{\evd}X} \\
\qquad =
     \ba[t]{@{}r@{~}l@{~}l@{~}l@{}}
              & s & x &\mapsto x \\
              & s & \effin{\op{get} \mapsto k} &\mapsto \op{state}~s~(k~s) \\
              & s & \effin{\op{set}~s' \mapsto k} &\mapsto \op{state}~s'~(k~unit)~\key{in} \\
     \ea \\
\key{let}~\op{index} : \forall \evd~X.\thunk{\effin{\id}\con{List}~X \to \effout{\evd}\con{List}~(\con{Pair}~\con{Nat}~X)} = \\
\qquad =
    \{                      \var{xs} \mapsto \op{state}~\con{zero}~(\op{map}~\thunk{x \mapsto \op{pair}~\op{next}!~x}~\var{xs})\}~\key{in} \\
\op{index}~\str{abc} \\
\el
\]
The $\op{map}$ function and $\op{state}$ handler are recursive, so are
defined using $\key{letrec}$, whereas the $\op{index}$ function is not
recursive so is defined with $\key{let}$. The type signatures are
adorned with explicit universal quantifiers and braces to denote that
they each define suspended computations. Pattern matching by equations
is represented by explicit pattern matching in the standard way. Each
wildcard pattern is represented with a fresh variable.




\subsection{Typing Rules}

The typing rules for Frank are given in Figure~\ref{fig:frank-typing}.
The inference judgement $\infersgs{m}{A}$ states that in type
environment $\Gamma$ with ambient ability $\sigs$, we can infer that
use $m$ has type $A$.
The checking judgement $\checksgs{A}{n}$ states that in type
environment $\Gamma$ with ambient ability $\sigs$, construction $n$
has type $A$.
The auxiliary judgement $\checksdefg{C}{e}$ states that in type
environment $\Gamma$, computation $e$ has type $C$.
The judgement $\matchesc{T}{r}{\sigs}{\Gamma}$
states that computation pattern $r$ of port type $T$ with ambient
ability $\sigs$ binds type environment $\Gamma$.
The judgement $\matchesv{A}{p}{\Gamma}$ states that value pattern $p$
of type $A$ binds type environment $\Gamma$.



The \textsc{Var} rule infers the type of a monomorphic variable $x$ by
looking it up in the environment; \textsc{PolyVar} does the same for a
polymorphic variable $f$, but also instantiates type variables and
effect variables through substitution $\theta$: the presentation is
declarative, so $\theta$ is unconstrained.
The \textsc{Command} rule infers the type of a command $c$ by looking
it up in the ambient ability, where the ports have the identity
adjustment and the peg has the ambient ability.

The \textsc{App} rule infers the type of an application $m~\many{n}$
under ambient ability $\sigs$. First it infers the type of $m$ of the
form $\thunk{\many{\effin{\adj}A} \to \effout{\sigs'}B}$. Then it
checks that $\sigs' = \sigs$ and that each argument $n_i$ matches the
inferred type in the ambient ability $\sigs$ extended with adjustment
$\adj_i$. If these checks succeed, then the inferred type for the
application is $B$.

The \textsc{Switch} rule allows us to treat a use as a
construction. The checking rules for data types (\textsc{Data}),
suspended computations (\textsc{Thunk}), polymorphic let
(\textsc{Let}), and mutual recursion (\textsc{LetRec}) recursively
check the subterms.

\paragraph{Notation}
We write $(M)_i$ for a list of zero or more copies of $M$ indexed by
$i$. Similarly, we write $(M)_{i,j}$ for a list of zero or more copies
of $M$ indexed by $i$ and $j$.
\medskip

A computation of type $\many{T \to}~ G$ is built by composing pattern
matching clauses of the form $\many{r} \mapsto n$ (\textsc{Comp}),
where $\many{r}$ is a sequence of computation patterns whose variables
are bound in $n$.
The side condition in the \textsc{Comp} rule requires that the
patterns in the clauses cover all possible values inhabiting the types
of the ports. Pattern elaboration (Section~\ref{sec:core}) yields an
algorithm for checking coverage.



Value patterns can be typed as computation patterns
(\textsc{P-Value}).
A request pattern $\effin{\handle{c~\many{p}}{z}}$ may be checked at
type $\effin{\adj}B'$ with ambient ability $\sigs$
(\textsc{P-Request}). The command $c$ must be in the adjustment
$\adj$. The continuation is a plain function so its port type has the
identity adjustment. The continuation's peg has the ambient ability
with $\adj$ applied. 
%
To check a computation pattern $\effin{x}$ we apply the adjustment to
the ambient ability (\textsc{P-CatchAll}).





\paragraph{Instantiating $\evd$}
In an earlier draft of this paper, the \textsc{PolyVar} rule was
restricted to always instantiate the implicit effect variable $\evd$
with the ambient ability $\sigs$. Correspondingly, data types were
restricted to being parameterised by at most one effect variable,
namely $\evd$.
The language resulting from these restrictions has the pleasant
property that effect variables need never be written at all.
However, we now feel that the restrictions are artificial, and having
multiple effect variables may be useful.
Given that the \textsc{App} rule already checks that ambients match up
exactly where needed, relaxing the \textsc{PolyVar} rule does no harm,
and now we can support data types parameterised by multiple effect
variables.

\paragraph{Subeffecting}
One strength of bidirectional type systems~\cite{PierceT00} is how
smoothly they extend to support subtyping rules.
Before adopting operators, we considered incorporating a subeffecting
judgement.
But, in the presence of operators, subeffecting does not seem
particularly helpful, as operators are invariant in their effects.

\section{Core Frank}
\label{sec:core}

We elaborate Frank into Core Frank, a language in which operators are
implemented through a combination of call-by-value functions, case
statements, and unary effect handlers.
Operators in Frank elaborate to $n$-ary functions over suspended
computations in Core Frank.
Shallow pattern matching on a single computation elaborates to unary
effect handling. Shallow pattern matching on a data type value
elaborates to case analysis. Deep pattern matching on multiple
computations elaborates to a tree of unary effect handlers and case
statements.

The abstract syntax of Core Frank is given in
Figure~\ref{fig:core-syntax}.
The only difference between Frank and Core Frank types is that a
computation type in Frank takes $n$ ports to a peg, whereas a
computation type in Core Frank takes $n$ value types to a peg.
The difference between the term syntaxes is more significant.
Polymorphic type variables are instantiated explicitly with type
arguments.
Constructions may be coerced to uses via a type annotation, which is
helpful for the semantics (Section~\ref{sec:semantics}), where
constructions are often substituted for variables.
In place of pattern-matching suspended computations, we have
$n$-argument lambda abstractions, case statements, and unary effect
handlers. The first two abstractions are standard; the third
eliminates a single effectful computation. Elimination of commands is
specified by command clauses which arise from request and catch-all
patterns in the source language. Elimination of return values is
specified by the single return clause, which arises from value
patterns in the source language. The adjustment annotation is
necessary for type checking.
%
As we no longer have operators, application is now plain $n$-ary
call-by-value function application.
\begin{figure}
\fighead{Types}
\begin{syntax}
\slab{value types}       &A, B     &::=& D~\many{R} \mid \thunk{C} \mid X \\
\slab{computation types} &C        &::=& \many{A \to}~G \\
\slab{pegs}              &G             &::=& \effbox{\sigs}A \\
[1ex]
\slab{type variables}    &Z        &::=& X \mid \effbox{\ev} \\
\slab{type arguments}    &R        &::=& A \mid \effbox{\sigs} \\
\slab{polytypes}         &P        &::=& \forall \many{Z}.A \\
[1ex]
\slab{abilities}    &\sigs  &::=&
  \nowt \mid \sigs, \sig~\many{R} \mid \ev \\
\slab{adjustments}  &\adj  &::=&
  \id \mid \adj + \sig~\many{R} \\
[1ex]
\slab{type environments}
                    &\Gamma        &::=& \cdot \mid \Gamma, x:A \mid f:P \\
\end{syntax}

\fighead{Terms}
\begin{syntax}
\slab{uses}         &m       &::=& x \mid f~\many{R} \mid c \mid m~s \mid (n : A) \\
\slab{constructions}&n       &::=& m \mid k~\many{n} \\
   &&\mid& \lambda \many{x}.n \\
   &&\mid& \key{case}~m~\key{of}~
             \many{k~\many{x} \mapsto n} \\
   &&\mid& \key{handle}^\adj_G~m~
             \ba[t]{@{}r@{}l@{}}
             \key{with}~& \many{\handle{c~\many{x}}{z} \mapsto n} \\
             \medvert~  &  x \mapsto n' \\
             \ea \\
   &&\mid&  \key{let}~f : P = n~\key{in}~n' \\
   &&\mid&  \key{letrec}~\many{f : P = \lambda \many{x}.n}~\key{in}~n' \\
[1ex]
\slab{spines}&s           &::=& \many{n} \\
\end{syntax}
\caption{Core Frank Abstract Syntax}
\label{fig:core-syntax}
\end{figure}

The Core Frank typing rules are given in Figure~\ref{fig:core-typing}.
\begin{figure*}
$\boxed{\infers{\Gamma}{\sigs}{m}{A}}$
\begin{mathpar}
\inferrule[Var]
  {x:A \in \Gamma}
  {\infersgs{x}{A}}

\inferrule[PolyVar]
  {f:P \in \Gamma} 
  {\infersgs{f~\many{R}}{P(\many{R})}}

\inferrule[Command]
  {c : \many{A \to}~ B \in \sigs}
  {\infersgs{c}{\thunk{\many{A \to}~ \effbox{\sigs}B}}}
\\
\inferrule[App]
  {\infersgs{m}{\thunk{\many{A \to}~ \effbox{\sigs'}B}} \\
   \sigs' = \sigs \\
   \many{\checks{\Gamma}{\sigs}{A}{n}}}
  {\infersgs{m~\many{n}}{B}}

\inferrule[Coerce]
  {\checksgs{A}{n}}
  {\infersgs{(n : A)}{A}}
\end{mathpar}

$\boxed{\checks{\Gamma}{\sigs}{A}{n}}$

\begin{mathpar}
\inferrule[Switch]
  {\infersgs{m}{A} \\ A = B}
  {\checksgs{B}{m}}

\inferrule[Data]
  {k~\many{A} \in D~\many{R} \\
   \many{\checksgs{A}{n}}}
  {\checksgs{D~\many{R}}{k~\many{n}}}

\inferrule[Fun]
  {\checks{\Gamma, \many{x}:\many{A}}{\sigs'}{B}{n}}
  {\checksgs{\thunk{\many{A \to}~ \effbox{\sigs'}B}}{\lambda \many{x}.n}}
\\
\inferrule[Case]
  {\infers{\Gamma}{\sigs}{m}{D~\many{R}} \\\\
   (\checks{\Gamma, \many{x}:\many{A}}{\sigs}{B}{n})_{k\,\many{A} \in D\,\many{R}}}
  {\checksgs{B}{\key{case}~ m ~\key{of}~
               (k~\many{x} \mapsto n)_k}}

\inferrule[Handle]
  {\infers{\Gamma}{\sigs \oplus \adj}{m}{A'} \\\\
   (\checks{\Gamma, \many{x}:\many{A}, z:B \to \effbox{\sigs \oplus \adj}A'}
    {\sigs}{B'}{n})_{c : \many{A \to}~ B ~\in~ \nowt \oplus \adj} \\
   \checks{\Gamma, x:A'}{\sigs}{B'}{n'}}
  {\checksgs{B'}{\key{handle}^\adj_{\effbox{\sigs}B'}~ m ~\key{with}~
               (\handle{c~\many{x}}{z} \mapsto n)_c \medvert
                x \mapsto n'}}

\inferrule[Let]
  {P = \many{\forall \many{Z}.A} \\
   \checksgs{A}{n} \\
   \checks{\Gamma, f : P}{\sigs}{B}{n'}}
  {\checksgs{B}{\key{let}~f : P = n~\key{in}~n'}}

\inferrule[LetRec]
  {\many{P} = \many{\forall \many{Z}.\thunk{C}} \\
   \many{\checks{\Gamma, \many{f : P}}{\sigs}{\thunk{C}}{\lambda \many{x}.n}} \\
   \checks{\Gamma, \many{f : P}}{\sigs}{B}{n'}}
  {\checksgs{B}{\key{letrec}~\many{f : P = \lambda \many{x}.n}~\key{in}~n'}}
\end{mathpar}

\caption{Core Frank Typing Rules}
\label{fig:core-typing}
\end{figure*}
They are mostly unsurprising given the corresponding Frank Typing
rules.
The \textsc{Handle} rule requires the adjustment $\adj$ for its
premisses (the source language builds this adjustment into ports).
It is also annotated with the result type.
The \textsc{Coerce} rule allows constructions to be treated as uses.
The type annotations in the \textsc{Handle} and \textsc{Coerce} rules
are used in the operational semantics.


\paragraph{Notation}
We extend our indexed list notation to allow indexing over data
constructors and commands. In typing rules, we follow the convention
that if a meta variable appears only inside such an indexed list then
it is implicitly indexed. For instance, the $n$ in the \textsc{Case} rule
depends on $k$, whereas $\sigs$ does not because it appears outside as
well as inside an indexed list.

\subsection{Elaboration}

We now describe the elaboration of Frank into Core Frank by way of a
translation $\sem{-}$.
%

We begin with the translation on types. Most cases are homomorphic,
that is, given by structurally recursive boilerplate, so we only give
the non-trivial cases.
In order to translate a computation type we supply the ambient ability
to each of the ports.
\[
\sem{\many{T \to}~\effout{\sigs}A} = \many{\sem{T}(\sem{\sigs}) \to}~\effout{\sem{\sigs}}\sem{A}
\]
Each port elaborates to a suspended computation type with effects
given by the ambient ability extended by the adjustment at the port.
\[
\sem{\effin{\adj}A}(\sigs) = \thunk{\effout{\sigs \oplus \sem{\adj}}\sem{A}}
\]

\newcommand{\PE}{\mathit{PE}}

\newcommand{\emptylist}{[]}
\newcommand{\cons}{\mathbin{::}}
\newcommand{\concat}{\mathbin{+\!+}}

\newcommand{\hdless}{\var{Headless}}
\newcommand{\true}{\text{true}}
\newcommand{\false}{\text{false}}
\newcommand{\Patterns}{\var{Patterns}}
\newcommand{\PatternTypes}{\var{PatternTypes}}
\newcommand{\As}{\var{As}}
\newcommand{\Bs}{\var{Bs}}
\newcommand{\Rs}{\var{Rs}}
\newcommand{\Qs}{\var{Qs}}
\newcommand{\ps}{\var{ps}}
\newcommand{\qs}{\var{qs}}
\newcommand{\rs}{\var{rs}}
\newcommand{\us}{\var{us}}
\newcommand{\vs}{\var{vs}}
\newcommand{\ws}{\var{ws}}
\newcommand{\xs}{\var{xs}}
\newcommand{\ys}{\var{ys}}

\newcommand{\proj}{\mathbin{@}}

The translation on terms depends on the type of the term so we specify
it as a translation on derivation trees.
As with the translation on types, we give only the non-trivial cases:
all of the other cases are homomorphic.
The translation on polymorphic variables converts implicit instantiation
into explicit type application.
\[
\left \llbracket
\inferrule
  {f:\forall \many{Z}.A \in \Gamma}
  {\infersgs{f}{\theta(A)}}
\right \rrbracket
=
\inferrule
  {f:\forall \many{Z}.\sem{A} \in \sem{\Gamma}}
  {\infersgs{f~\many{R}}{\sem{A}[\many{R}/\many{Z}]}},
~~  \\
\]
where $R_i = \sem{\theta(Z_i)}$ and $A[\many{R}/\many{Z}]$ denotes the
simultaneous substitution of each type argument $R_i$ for type
variable $Z_i$ in value type $A$.
In the remaining non-trivial cases, we save space by writing only the
judgement at the root of a derivation and by writing only the term
when referring to a descendent of the root.

The crux of the translation is the elaboration of pattern matching for
computations. Computations can occur either in suspended computations
or mutually recursive definitions. In order to translate a computation
we supply the ambient ability.
\[
\bl
\sem{\checksgs{C}{\thunk{e}}} = \sem{e}(\sem{\sigs}) \\
\sem{\checksgs{B}{\key{letrec}~\many{f : P = e}~\key{in}~n}} = \\
\qquad
  \sem{\Gamma} \sigentails{\sem{\sigs}} \key{letrec}~\many{f : \sem{P} = \sem{e}(\sem{\sigs})}~\key{in}~\sem{n} \\
\el
\]
The translation of a computation is then defined as follows
\[
\bl
\sem{\checksdefg{(T_j \to)_j~ G}{{((r_{i,j})_j \mapsto n_i)_i}}}(\sigs) = \\
\qquad
 \sem{\Gamma} \sigentails{\sigs}
   \bl
   \lambda (x_j)_j.
     \PE((x_j)_j, (T_j)_j, ((r_{i,j})_j \mapsto \sem{n_i})_i, G) \\
   \quad~:~\thunk{(\sem{T_j} \to)_j~ \sem{G}}
   \el \\
\el
\]
where each $x_j$ is fresh and $\PE(\many{x}, \many{Q}, \many{u}, G)$
is a function that takes a list of variables $\many{x}$ to eliminate,
a list of pattern types $\many{Q}$, a pattern matching matrix
$\many{u}$, and a result type $G$, and yields a Core Frank term.
Pattern types ($Q$) are either value types ($A$), port types ($T$), or
inscrutable types ($\bullet$). We use the latter to avoid trying to
reconstruct continuation types.
%
%
A pattern matching matrix $\many{u}$ is a list of pattern matching
clauses, where the body $n$ of each clause $u = \many{r} \mapsto n$ is
a Core Frank construction instead of a source Frank construction.

The pattern matching elaboration function $\PE$ is defined in
Figure~\ref{fig:pattern-matching-elaboration}. For this purpose we
find it convenient to use functional programming list notation. We
write $\emptylist$ for the empty list, $v \cons \vs$ for the list
obtained by prepending element $v$ to the beginning of the list $\vs$,
$[v]$ as shorthand for $v \cons \emptylist$, and $\vs \concat \ws$ for
the list obtained by appending list $\ws$ to the end of list $\vs$.
\begin{figure*}
\begin{equations}
\PE(x \cons \xs, D\,\Rs \cons \Qs, \us, G) &=&
    \bl
    \PE(\xs, \Qs, \us \proj x, G)  \quad \text{if } \hdless(\us)
    \el \\
\PE(x \cons \xs, D\,\Rs \cons \Qs, \us, G) &=&
    \bl
    \key{case}~x~\key{of}\\
    \quad
      (k_i~\ys_i \mapsto \PE(\ys_i \concat \xs, \PatternTypes(D\,\Rs, k_i) \concat \Qs, \us \proj k_i~\ys_i, G))_i\\
    \text{where} \\
    \quad (k_i~\ys_i)_i = \Patterns(D) \\
    \el\\
\PE(x \cons \xs, \effin{\adj}A \cons \Qs, \us, G) &=&
    \bl
    \key{handle}^{\sem{\adj}}_{\sem{G}}~x!~\key{with} \\
    \quad
    \bl
      (\effin{\handle{c_i~\ys_i}{z_i}} \mapsto
         \PE(z_i \cons \ys_i \concat \xs, \PatternTypes(\adj, c_i) \concat \Qs, \us \proj \effin{\handle{c_i~\ys}{z_i}}, G))_i \\
      w \mapsto \PE(w \cons \xs, A \cons \Qs, \us, G) \\
    \el \\
    \text{where} \\
    \quad w \cons (\effin{\handle{c_i~\ys_i}{z_i}})_i = \Patterns(\adj) \\
    \el \\
\PE(x \cons \xs, Q \cons \Qs, \us, G) &=& \PE(\xs, \Qs, \us \proj x, G) \\
\PE(\emptylist, \emptylist, (\emptylist \mapsto n) \cons \us, G) &=& n \\
\end{equations}

\begin{equations}
  \hdless(\emptylist) &=& \true\\
  \hdless(u \cons \us) &=& \hdless(u) \wedge \hdless(us)\\[1ex]

  \hdless(x \cons \rs \mapsto n) &=& \true\\
  \hdless(\effin{x} \cons \rs \mapsto n) &=& \true\\
  \hdless(r \cons \rs \mapsto n) &=& \false\\
\end{equations}

\begin{equations}
   \emptylist \proj r &=& \emptylist \\
(u \cons \us) \proj r &=& (u \proj r) \cons (\us \proj r) \\[1ex]
(k~\ps' \cons \rs \mapsto n) \proj k~\ps &=&
        [\ps \concat \rs \mapsto \key{let}~\ps' = \ps~\key{in}~n] \\
(k'~\ps' \cons \rs \mapsto n) \proj k~\ps &=& [], \quad \text{if }k \neq k'\\
    (x \cons \rs \mapsto n) \proj k~\ps &=&
        [\ps \concat \rs \mapsto \key{let}~x = k~\ps~\key{in}~n] \\
(\effin{x} \cons \rs \mapsto n) \proj k~\ps &=& [ps \concat \rs \mapsto
          \key{let}~x = \thunk{k~\ps}~\key{in}~n]\\
     (r \cons \rs \mapsto n) \proj k~\ps &=& \emptylist \\[1ex]
%
%

(\effin{\handle{c~\ps'}{q'}} \cons \rs \mapsto n) \proj \effin{\handle{c~\ps}{q}} &=&
  [q \cons \ps \concat \rs \mapsto \key{let}~(q' \cons \ps') = (q \cons \ps)~\key{in}~n] \\
(\effin{x} \cons \rs \mapsto n) \proj \effin{\handle{c~\ps}{q}} &=&
  [q \cons \ps \concat \rs \mapsto \key{let}~x = \thunk{q~(c~\ps)}~\key{in}~n] \\
(r \cons \rs \mapsto n) \proj \effin{\handle{c~\ps}{q}} &=& \emptylist \\[1ex]
(y \cons \rs \mapsto n) \proj x &=& [\rs \mapsto \key{let}~y = x~\key{in}~n] \\
(\effin{y} \cons \rs \mapsto n) \proj x &=&
  [\rs \mapsto \key{let}~y = \thunk{x}~\key{in}~n] \\
\end{equations}

\[
\ba{@{}r@{~}c@{~}l@{\quad}l@{}}
\Patterns(D) &=& (k~\xs_k)_{k \in D},  &\text{each }\xs_k \text{ fresh} \\
\Patterns(I) &=& (\effin{\handle{c~\xs_c}{y_c}})_{c \in I}, &\text{each }\xs_c, y_c \text{ fresh} \\
\ea
\qquad
\begin{eqs}
\Patterns(\id) &=& [w], \quad w \text{ fresh} \\
\Patterns(\Delta + I~\Rs) &=& \Patterns(\Delta) \concat \Patterns(I) \\
\end{eqs}
\]

\[
\ba{@{}r@{~}c@{~}l@{\quad}l@{}}
\PatternTypes(D\,\Rs, k) &=& \As, & \text{where }\dcon(D\,\Rs, k) = \As \\
\PatternTypes(I\,\Rs, c) &=& \bullet \cons \As,   & \text{where }\icom(I\,\Rs, c) = \As \to B \\
\PatternTypes(I\,\Rs, c) &=& [], & \text{if }\icom(I\,\Rs, c)\text{ undefined} \\
\ea
\qquad
\begin{eqs}
\PatternTypes(\id) &=& \emptylist \\
\PatternTypes(\Delta + I~\Rs, c) &=& \\
\multicolumn{3}{@{\qquad}l@{}}{\PatternTypes(\Delta, c) \concat \PatternTypes(I~\Rs, c)} \\
\end{eqs}
\]

\caption{Pattern Matching Elaboration}
\label{fig:pattern-matching-elaboration}
\end{figure*}
There are four cases for $\PE$. If the head pattern type is a data type,
then it generates a case split. If the head pattern is a port type
then it generates a handler. If the head pattern is some other pattern
type (a suspended computation type or a type variable) then neither
eliminator is produced. If the lists are empty then the body of the
head clause is returned.

\paragraph{Sequencing Computations}
We write $\key{let}~x=n~\key{in}~n'$ as syntactic sugar for
$\var{on}~n~\thunk{x \mapsto n'}$, where $\var{on}$ is as in
Section~\ref{sec:examples}:
\[
\bstack
\key{let}~x=n~\key{in}~n' \equiv \\
\quad \key{let}~
  (\var{on} :
    \forall \evd~X~Y.
       \thunk{\effin{\id}X \to \effin{\id}\thunk{\effin{\id}X \to \effout{\evd}Y} \to \effout{\evd}Y}) = \\
  \qquad \thunk{x~f~\mapsto~f~x}~\key{in}~\var{on}~n~\thunk{x \mapsto n'}
\estack
\]
This sugar differs from the polymorphic let construct in two ways: 1)
it has no type annotation on $n$, and 2) $x$ is monomorphic in $n'$.
\medskip

We make use of several auxiliary functions. The $\Patterns$ function
returns a complete list of patterns associated with the supplied data
type or interface. The $\PatternTypes$ function takes a data type and
constructor or interface and command, and returns a list of types of
the components of the constructor or command. The operation $\us \proj
r$ projects out a new pattern matching matrix from $\us$ filtered by
matching the pattern $r$ against the first column of $\us$. We make
use of the obvious generalisations of let binding for binding multiple
constructions and for rebinding patterns.

\paragraph{Example}
To illustrate how operators are elaborated into Core Frank, we give
the Core Frank representation of the \verb!pipe! multihandler defined
in Section~\ref{subsec:pipe}.

\[
\bl
\key{letrec}~
  \op{pipe} : \\
    \quad \forall \evd~X~Y. \{\bl
              \thunk{\effbox{\evd, \inter{Abort}, \inter{Send}~X}\con{Unit}} \to \\
              \quad \thunk{\effbox{\evd, \inter{Abort}, \inter{Receive}~X}Y} \to
                \effbox{\evd, \inter{Abort}}Y\} \\
              \el \\
\quad\qquad =
  \lambda x\,y.
    \bl
    \key{handle}^{\id + \inter{Send}\,X}_{\effbox{\evd, \inter{Abort}}Y}\,x!~\key{with} \\
       \ba{@{}r@{\quad}l@{}}
       &\op{\effin{\handle{\op{send}~x}{s}}} \mapsto \\
       &\quad
          \bl
          \key{handle}^{\id + \inter{Receive}\,X}_{\effbox{\evd, \inter{Abort}}Y}\,y!~\key{with} \\
            \ba{@{}r@{\quad}l@{~}l@{}}
                     &\op{\effin{\handle{\op{receive}}{r}}} &\mapsto \op{pipe}~(s~\op{unit})~(r~x) \\
            &y                                     &\mapsto y \\
            \ea \\
          \el \\
          &x \mapsto
             \bl
             \key{case}~x~\key{of}~\con{unit} \mapsto \\
             \quad \bl
                   \key{handle}^{\id + \inter{Receive}\,X}_{\effbox{\evd, \inter{Abort}}Y}\,y!~\key{with} \\
                      \ba{@{}r@{\quad}l@{~}l@{}}
                               &\op{\effin{\handle{\op{receive}}{r}}} &\mapsto \op{abort}! \\
                      &y                                     &\mapsto y \\
                      \ea \\
                   \el \\
             \el \\
       \ea \\
   \el \\
\el
\]
The ports are handled left-to-right. The producer is handled first. A
different handler for the consumer is invoked depending on whether the
producer performs a $\op{send}$ command or returns a value.

Notice that the type pollution problem described in
Section~\ref{subsec:pipe} is manifested in the translation to Core
Frank: the types of the arguments are unable to distinguish the handled
interfaces from those that just happen to be in the ambient ability.

Our pattern matching elaboration procedure is rather direct, but is
not at all optimised for efficiency. We believe it should be
reasonably straightforward to adapt standard techniques
(e.g.~\cite{Maranget08}) to implement pattern matching more
efficiently.
However, some care is needed as pattern matching compilation
algorithms often reorder columns as an optimisation. Column reordering
is not in general a valid optimisation in Frank. This is because
commands in the ambient ability, but not in the argument adjustments,
are implicitly forwarded, and the order in which they are forwarded is
left-to-right.
(Precise forwarding behaviour becomes apparent when we
combine pattern elaboration with the operational semantics for Core
Frank in Section~\ref{sec:semantics}.)

\subsection{Incomplete and Ambiguous Pattern Matching as Effects}
\label{sec:coverage-redundancy}

The function $\PE$ provides a straightforward means for checking
coverage and redundancy of pattern matching. Incomplete coverage can
occur iff $\PE$ is invoked on three empty lists $\PE(\emptylist,
\emptylist, \emptylist, G)$, which means $\PE$ is undefined on its
input. Redundancy occurs iff the final clause defining $\PE$ in
Figure~\ref{fig:pattern-matching-elaboration} is invoked in a
situation in which $\us$ is non-empty.
As an extension to Frank, we could allow incomplete and ambiguous
pattern matching mediated by effects.
We discuss this possibility further in
section~\ref{sec:future}.

\begin{figure*}

\begin{equations}
(\textrm{use values}) \quad v &::=& x \mid f~\many{R} \mid c \mid (w : A) \\
(\textrm{construction values}) \quad w &::=& v \mid k~\many{w} \mid \lambda \many{x}.n \\
(\textrm{evaluation contexts})\quad \EC
  &::= & [~] \mid \EC~\many{n} \mid v~(\many{w},\EC,\many{n})
             \mid (\EC : A)
             \mid k~(\many{w},\EC,\many{n})
             \mid \key{case}~\EC~\key{of}~(k\,\many{x_k} \mapsto n_k)_k \\
  &\mid& \key{handle}^\adj_G~ \EC ~\key{with}~
            (\handle{c~\many{x_c}}{z_c} \mapsto n_c)_c \medvert
            x \mapsto n' \\
  &\mid& \key{let}~(f : P) = \EC~\key{in}~n \\
\end{equations}%

\begin{equations}
(\lambda \many{x}.n : \thunk{\many{A \to}~\effbox{\sigs}B})~\many{w} &\reducesto& (n[\many{(w : A)} / \many{x}] : B) \\
\key{case}~(k'~\many{w} : D\,\many{R})~\key{of}~(k~\many{x_k} \mapsto n_k)_k
  &\reducesto& n_{k'}[\many{(w : A)} / \many{x_{k'}}],
     \quad \many{A} = \dcon(D\,\many{R}, k') \\
\key{handle}^\adj_G~v~\key{with}~(\handle{c~\many{x_c}}{z_c} \mapsto n_c)_c \medvert x \mapsto n'
  &\reducesto& n'[v / x] \\
\key{handle}^\adj_G~\EC[c'~\many{w}]~\key{with}~(\handle{c~\many{x_c}}{z_c} \mapsto n_c)_c \medvert x \mapsto n'
  &\reducesto& n_{c'}[\many{(w : A)} / \many{x_{c'}}, (\lambda y.\EC[y] : \thunk{B \to G}) / z_{c'}],
  ~~ c' \notin \var{HC}(\EC) \text{ and } c':\many{A \to}~B \in \adj \\
\key{let}~f : P = w~\key{in}~n
  &\reducesto&
    n[(w : P) / f] \\
\key{letrec}~\many{f : P = \lambda \many{x}.n}~\key{in}~n'
  &\reducesto&
    n'[\many{(\lambda \many{x}.\key{letrec}~\many{f : P = \lambda \many{x}.n}~\key{in}~n : P) / f}] \\
\\
(v : A) &\reducesto& v \\
\\
\EC[n] &\reducesto& \EC[n'], \quad \text{if }n \reducesto n' \\
\end{equations}%

\caption{Small-Step Operational Semantics for Core Frank}
\label{fig:semantics}
\end{figure*}

\section{Small-Step Operational Semantics}
\label{sec:semantics}

We give a small step operational semantics for Core Frank inspired by
Kammar et al.'s semantics for the effect handler calculus
\lameff~\cite{KammarLO13}.
The main differences between their semantics and ours arise from
differences in the calculi. Whereas \lameff is call-by-push-value,
Core Frank is $n$-ary call-by-value, which means Core Frank has many
more kinds of evaluation context.
%
A more substantive difference is that handlers in \lameff are deep
(the continuation reinvokes the handler), whereas handlers in Frank
are shallow (the continuation does not reinvoke the handler).
%

The semantics is given in Figure~\ref{fig:semantics}.  All of the
rules except the ones for handlers are standard $\beta$-reductions
(modulo some typing noise due to bidirectional typing).
We write $n[m / x]$ for $n$ with $m$ substituted for $x$ and
$n[\many{m} / \many{x}]$ for $n$ with each $m_i$ simultaneously
substituted for each $x_i$.
Similarly, we write $n[(n' : P)/f]$ for $n$ with $(n' : P(\many{R}))$
substituted for $f~\many{R}$ and the corresponding generalisation for
simultaneous substitution (writing $P(\many{R})$ for
$A[\many{R}/\many{Z}]$ where $P = \forall \many{Z}.A$).
Values are handled by substituting the value into the handler's return
clause. Commands are handled by capturing the continuation up to the
current handler and dispatching to the appropriate clause for the
command. We write $\var{HC}(\EC)$ for the set of commands handled by
evaluation context $\EC$. Formally $\var{HC}(-)$ is given by the
homomorphic extension of the following equations.
\[
\bl
\var{HC}([~]) = \emptyset \\
\var{HC}(\key{handle}^\adj_G~ \EC ~\key{with}~
            (\handle{c_i~\many{x_{c_i}}}{z_{c_i}} \mapsto n_{c_i})_i \medvert
            x \mapsto n') =\quad \\
\hfill \var{HC}(\EC) \cup \{c_i\}_i \\
\el
\]
The side condition on the command rule ensures that command $c'$ is
handled by the nearest enclosing handler that has a clause for
handling $c'$.
A more intensional way to achieve the same behaviour is to explicitly
forward unhandled commands using an additional rule~\cite{KammarLO13}.



Reduction preserves typing.
\begin{theorem}[Subject Reduction]
~
\begin{itemize}
\item If $\infersgs{m}{A}$ and $m \reducesto m'$ then $\infersgs{m'}{A}$.
\item If $\checksgs{A}{n}$ and $n \reducesto n'$ then $\checksgs{A}{n'}$.
\end{itemize}
\end{theorem}

There are two ways in which reduction can stop: it may yield a value,
or it may encounter an unhandled command instance (if the ambient
ability is non-empty). We capture both possibilities with a notion of
normal form, which we use to define type soundness.

\begin{definition}[Normal Forms]
If $\checksgs{A}{n}$ then we say that $n$ is normal with respect to
$\sigs$ if it is either a value $w$ or of the form $\EC[c~\many{w}]$
where $c : \many{A} \to B \in \sigs$ and $c \notin \var{HC}(\EC)$.
\end{definition}

\begin{theorem}[Type Soundness]
~\\ If $\checks{\cdot}{\sigs}{A}{n}$ then either $n$ is normal with
  respect to $\sigs$ or there exists $\checks{\cdot}{\sigs}{A}{n'}$
  such that $n \reducesto n'$. (In particular, if $\sigs = \nowt$ then
  either $n$ is a value or there exists $\checks{\cdot}{\sigs}{A}{n'}$
  such that $n \reducesto n'$.)
\end{theorem}



\section{Computations as Data}
\label{sec:compdata}

So far, our example data types have been entirely first order, but our
type system admits data types which abstract over abilities exactly to
facilitate the storage of suspended computations in a helpfully
parameterised way. When might we want to do that? Let us develop an
example, motivated by Shivers and Turon's treatment of \emph{modular
  rollback} in parsing~\cite{DBLP:conf/icfp/ShiversT11}.

Consider a high-level interface to an input stream of characters
with one-step lookahead. A parser may \verb!peek! at the next
input character without removing it, and \verb!accept! that character
once its role is clear.
\begin{verbatim}
interface LookAhead = peek : Char | accept : Unit
\end{verbatim}
We might seek to implement \verb!LookAhead! on top of regular
\verb!Console! input, specified thus:
\begin{verbatim}
interface Console = inch : Char
                  | ouch : Char -> Unit
\end{verbatim}
where an input of \verb!'\b'! indicates that the backspace key has
been struck.
The appropriate behaviour on receipt of backspace is to unwind the
parsing process to the point where the previous character was first
used, then await an alternative character.  To achieve that unwinding,
we need to keep a \emph{log}, documenting what the parser was doing
when the console actions happened.
\begin{verbatim}
data Log X
  = start  {X}
  | inched (Log X) {Char -> X}
  | ouched (Log X)
\end{verbatim}
Note that although \verb!Log! is not explicitly parameterised by an
ability it must be implicitly parameterised as it stores implicitly
effect polymorphic suspended computations.
The above definition is shorthand for the following.


\smallskip
{\tt
\begin{tabular}{@{\hspace{-3ex}}l@{}}
data Log [$\evd$] X \\
~~= start  \texttt{\char`\{}[$\evd$]X\texttt{\char`\}} \\
~~| inched (Log [$\evd$] X) \texttt{\char`\{}Char -> [$\evd$]X\texttt{\char`\}} \\
~~| ouched (Log [$\evd$] X)
\end{tabular}}
\smallskip

%
\noindent
As discussed in Section~\ref{sec:frank}, the general rule is that if
the body of a data type (or interface) definition includes the
implicit effect variable then the first parameter of the definition is
also the implicit effect variable.
%
%
Initially a log contains a parser computation (\verb!start!). When a
character is input (\verb!inched!) the log is augmented with the
continuation of the parser, which depends on the character read,
allowing the continuation to be replayed if the input character
changes. When a character is output (\verb!ouched!) there is no need
to store the continuation as the return type of an output is
\verb!Unit! and hence cannot affect the behaviour of the parser.

Modular rollback can now be implemented as a handler informed by a log
and a one character buffer.
\begin{verbatim}
data Buffer = empty | hold Char
\end{verbatim}

The parser process being handled should
also be free to reject its input by aborting, at which point the
handler should reject the character which caused the rejection.
\begin{verbatim}
input :  Log [LookAhead, Abort, Console] X ->
         Buffer ->
         <LookAhead, Abort>X ->
         [Console]X
input _ _        x               = x
input l (hold c) <peek -> k>     =
  input l (hold c) (k c)
input l (hold c) <accept -> k>   =
  ouch c; input (ouched l) empty (k unit)
input l empty    <accept -> k>   =
  input l empty (k unit)
input l empty    <peek -> k>     =
  on inch!
     { '\b' -> rollback l
     | c    -> input (inched l k) (hold c) (k c) }
input l _        <aborting -> k> = rollback l
\end{verbatim}
Note that the \verb!Log! type's ability has been instantiated with
exactly the same ambient ability as is offered at the port in which
the parser plugs. Correspondingly, it is clear that the parser's
continuations may be stored, and under which conditions those
stored continuations can be invoked, when we \verb!rollback!.
\begin{verbatim}
rollback : Log [LookAhead, Abort, Console] X ->
           [Console]X
rollback (start p)    = parse p
rollback (ouched l)   = map ouch "\b \b";
                        rollback l
rollback (inched l k) = input l empty (k peek!)

parse : {[LookAhead, Abort, Console]X} ->
          [Console]X
parse p = input (start p) empty p!
\end{verbatim}
To undo an \verb!ouch!, we send a backspace, a blank, and another
backspace, erasing the character. To undo the \verb!inch! caused by a
`first \verb!peek!', we empty the buffer and reinvoke the old
continuation after a new \verb!peek!.

Here is a basic parser that accepts a sequence of zeros terminated by
a space, returning the total length of the sequence on success, and
aborting on any other input.
\begin{verbatim}
zeros : Int -> [LookAhead, Abort]Int
zeros n = on peek! { '0' -> accept!; zeros (n+1)
                   | ' ' -> accept!; n
                   | c   -> abort!}
\end{verbatim}

In order to implement actual console input and output the
\verb!Console! interface is handled specially at the top-level using a
built-in handler that interprets \verb!inch! and \verb!ouch! as actual
character input and output.
The entry point for a Frank program is a nullary \verb!main! operator.
\begin{verbatim}
main : [Console]Int
main! = parse (zeros 0)
\end{verbatim}
The ability of \verb!main! is the external ability of the whole
program. We can use it to configure the runtime to the execution
context: is it a terminal? is it a phone? is it a browser? What will
the user \emph{let} us do?
Currently, Frank supports a limited range of built-in top-level
interfaces, but one can imagine adding many more and in particular
connecting them to external APIs.

While the \verb!Log! type does what is required of it, this example
does expose a shortcoming of Frank as currently specified: we have
no means to prevent the parser process from accessing \verb!Console!
commands, because our adjustments can add and shadow interfaces but
not remove them. If we permitted `negative' adjustments, we could
give the preferable types
\begin{verbatim}
input :  Log [LookAhead, Abort] X -> Buffer ->
         <LookAhead, Abort, -Console>X ->
         [Console]X

rollback : Log [LookAhead, Abort] X -> [Console]X

parse : {[LookAhead, Abort]X} -> [Console]X
\end{verbatim}
At time of writing, it is clear how to make negative adjustments act
on a concrete ability, but less clear what their impact is on effect
polymorphism---a topic of active investigation.

\section{Implementation}
\label{sec:implementation}

The second author has been plotting Frank since at least
2007~\cite{McBride07}. In 2012, he implemented a prototype for a
previous version of Frank~\cite{McBride12}. Since then the design has
evolved. A significant change is the introduction of operators that
handle multiple computations simultaneously.
More importantly, a number of flaws in the original design have been
ironed out as a result of formalising the type system and semantics.

We have now implemented a prototype of Frank in Haskell that matches
the design described in the current paper~\cite{McLaughlin16}.
In order to rapidly build a prototype, we consciously decided to
take advantage of a number of existing technologies.
The current prototype takes advantage of the indentation
sensitive parsing framework of \citeauthor{AdamsA14}~\cite{AdamsA14},
the ``type-inference-in-context'' technique of
\citeauthor{GundryMM10}~\cite{GundryMM10}, and the existing
implementation of Shonky~\cite{McBride16}.

Much like Haskell, in order to aid readability, the concrete syntax of
Frank is indentation sensitive (though we do not explicitly spell out
the details in the paper).
In order to implement indentation sensitivity,
\citeauthor{AdamsA14}~\cite{AdamsA14} introduce an extension to
parsing frameworks based on parsing expression grammars. Such grammars
provide a formal basis for the Parsec~\cite{LeijenParsec} and
Trifecta~\cite{KmettTrifecta} parser combinator Haskell libraries. In
contrast to the ad hoc methods typically employed by many indentation
sensitive languages (including Haskell and Idris~\cite{BradyIdris13}),
\citeauthor{AdamsA14}'s extension has a formal semantics. Frank's
parser is written using Trifecta with the indentation sensitive
extension, which greatly simplifies the handling of indentation by
separating it as much as possible from the parsing process.

For bidirectional typechecking, our prototype uses
\citeauthor{GundryMM10}'s ``type-inference-in-context''
technique~\cite{GundryMM10} for implementing type inference and
unification (Gundry's thesis~\cite{GundryThesis} contains a more
detailed and up-to-date account).
The key insight is to keep track in the context not just of term
variables but also unification variables. The context enforces an
invariant that later bindings may only depend on earlier bindings.
The technique has been shown to scale to the dependently typed
setting~\cite{GundryThesis}.

The back-end of the prototype diverges from the formalism described in
the paper. Instead of targeting a core calculus, Frank is translated
into Shonky~\cite{McBride16}, which amounts to an untyped version of
Frank. Shonky executes code directly through an abstract machine much
like that of Hillerstr\"om and Lindley~\cite{HillerstromL16}.

\section{Related Work}
\label{sec:related}

We have discussed much of the related work throughout the paper. Here
we briefly mention some other related work.

\paragraph{Efficient Effect Handler Implementations}

A natural implementation for handlers is to use \emph{free
  monads}~\cite{KammarLO13}. Swierstra~\cite{Swierstra08} illustrates
how to write effectful programs with free monads in Haskell, taking
advantage of type-classes to provide a certain amount of modularity.
However, using free monads directly can be quite
inefficient~\cite{KammarLO13}.

Wu and Schrijvers~\cite{WuS15} show how to obtain a particularly
efficient implementation of deep handlers taking advantage of
fusion. Their work explains how Kammar et al.~\cite{KammarLO13} implemented
efficient handler code in Haskell.
Kiselyov and Ishii~\cite{KiselyovI15} optimise their shallow effect
handlers implementation, which is based on free monads, by taking
advantage of an efficient representation of sequences of monadic
operations~\cite{PloegK14}.
The experimental multicore extension to OCaml~\cite{Dolan2015} extends
OCaml with effect handlers motivated by a desire to abstract over
scheduling strategies. It does not include an effect system. It does
provide an efficient implementation by optimising for the common case
in which continuations are invoked at most once (the typical case for
a scheduler). The implementation uses the stack to represent
continuations and as the continuation is used at most once there is no
need to copy the stack.
Koka~\cite{Leijen17} takes advantage of a selective CPS translation to
improve the efficiency of generated JavaScript code.

\paragraph{Layered Monads and Monadic Reflection}
%

Filinski's work on monadic reflection~\cite{Filinski10} and layered
monads~\cite{Filinski99} is closely related to effect
handlers. Monadic reflection supports a similar style of composing
effects. The key difference is that monadic reflection interprets
monadic computations in terms of other monadic computations, rather
than abstracting over and interpreting operations

Swamy et al.~\cite{SwamyGLH11} add support for monads in ML, providing
direct-style effectful programming for a strict language. Unlike
Frank, their system is based on monad transformers rather than effect
handlers.

Schrijvers et al.~\cite{SchrijversPWJ16} compare the expressiveness of
effect handlers and monad transformers in the context of Haskell.
Forster et al.~\cite{ForsterKLP16} compare effect handlers with
monadic reflection and delimited control in a more abstract setting.

\paragraph{Variations and Applications}
Lindley~\cite{Lindley14} investigates an adaptation of effect handlers
to more restrictive forms of computation based on
idioms~\cite{McbrideP08} and arrows~\cite{Hughes04}.
Wu et al.~\cite{WuSH14} study scoped effect handlers. They attempt to
tackle the problem of how to modularly weave an effect handler through
a computation whose commands may themselves be parameterised by other
computations.
Kiselyov and Ishii~\cite{KiselyovI15} provide solutions to particular
instances of this problem.
Schrijvers et al.~\cite{SchrijversWDD14} apply effect handlers to
logic programming.


\section{Future Work}
\label{sec:future}

We have further progress to make on many fronts, theoretical and
practical.


\paragraph{Direct Semantics}
The semantics of Frank presented in this paper is via a translation to
Core Frank, a fairly standard call-by-value language extended with
algebraic effects and unary effect handlers.
This has the advantage of making it clear how the semantics of Frank
relates to the semantics of other languages with algebraic effects and
effect handlers.
Nevertheless, we believe there are good reasons to explore a more
direct semantics.
A direct semantics may offer a more efficient implementation or more
parsimonious generated code (witness the translation of \verb!pipe!
into Core Frank in Section~\ref{sec:core}).
Perhaps a more compelling motivation is that multihandlers appear to
share some features of other expressive programming abstractions such
as multiple dispatch and join patterns~\cite{FournetG96}, and it would
be interesting to better understand how multihandlers relate to these
abstractions.

\paragraph{Verbs versus Nouns}
Our rigid choice that names stand for values means that nullary
operators need \verb|!| to be invoked. They tend to be much more
frequently found in the doing than the being, so it might be
prettier to let a name like \verb!jump! stand for the `intransitive
verb', and write \verb!{jump}! for the `noun'. Similarly, there is
considerable scope for supporting conveniences such as giving
functional computations by partial application whenever it is
unambiguous.

\paragraph{Data as Computations}
Frank currently provides both interfaces and data types. However,
given an empty type $0$, we can simulate data types using interfaces
(and data constructors using commands).
Data constructors can be seen as exceptions, that is, commands that do
not return a value.
For instance, we can encode lists as follows:
\begin{verbatim}
interface ListI X = nil : 0
                  | cons : X -> ListI X -> 0
\end{verbatim}
The type of lists with elements of type \verb!X! is \verb![ListI X]0!.
We can then simulate \verb!map! as follows:
\begin{verbatim}
mapI : {X -> Y} -> [ListI X]0 -> [ListI Y]0
mapI f <nil       -> _> = nil!
mapI f <cons x xs -> _> = cons (f x) (mapI f xs)
\end{verbatim}
Note that the pattern matching clauses are complete because the return
type is uninhabited.

Given that computations denote free monads (i.e, trees) and data types
also denote trees, it is hardly surprising that there is a
correspondence here.
Indeed Atkey's algebraic account of type checking and
elaboration~\cite{Atkey15} makes effective use of this correspondence.
We would like to study abstractions for more seamlessly translating
back and forth between computations and data.

\paragraph{Failure and Choice in Pattern Matching}
As discussed in Section~\ref{sec:coverage-redundancy} we could extend
Frank to realise incomplete or ambiguous patterns as
effects.
Pattern matching can fail (if the patterns are incomplete) or it can
succeed in multiple ways (if the patterns are redundant). Thus in
general pattern matching yields a searchable solution space.
We can mediate failure and choice as effects, separating what it is to
\emph{be} a solution from the strategy used to \emph{find} one.
Concretely, we envisage the programmer writing custom failure and
choice handlers for navigating the search space.
Wu, Schrijvers and Hinze~\cite{WuSH14} have shown the modularity and
flexibility of effect handlers in managing backtracking computations:
the design challenge is to deploy that power in the pattern language
as well as in the expression language.

\paragraph{Scaling by Naming}
What if we want to have multiple \verb!State! components? One
approach, adopted by Brady~\cite{Brady13}, is to rename them
apart: when we declare the \verb!State! interface, we acquire
also the \verb!Foo.State! interface with operations \verb|Foo.get|
and \verb|Foo.set|, for any \verb|Foo|. We would then need to
specialise \verb!State!ful operators to a given \verb|Foo|,
and perhaps to generate fresh \verb|Foo|s dynamically.

\paragraph{Dynamic Effects}
An important effect that we cannot implement directly in Frank as it
stands is dynamic allocation of ML-style references. One difficulty is
that the \verb|new| command which allocates a new reference cell has a
polymorphic type \verb!forall X.new : Ref X!. But even if we restrict
ourselves to a single type, it is still unclear how to safely
represent the \verb!Ref! data type. Eff works around the problem using
a special notion of resource~\cite{BauerP15}. We would like to explore
adding resources or a similar abstraction to Frank.




\paragraph{Negative Adjustments}
Currently the only non-trivial adjustments are positive: they add
interfaces to an ability. As mentioned in Section~\ref{sec:compdata},
we would like to add support for negative adjustments that remove
interfaces from an ability. An extreme case of a negative adjustment
is a \emph{nugatory} adjustment that removes all interfaces from any
ability to yield the pure ability.
Whereas positive adjustments can be simulated with more conventional
effect type systems, at the cost of some precision in types, it is not
clear to us whether negative adjustments can be.

\paragraph{Controlled Snooping}
We might consider allowing handlers to trap some or even all commands
generically, just as long as their ports make this possibility
clear. Secret interception of commands remains anathema.

\paragraph{Indexed Interfaces}
Often, an interaction with the environment has some sort of state,
affecting which commands are appropriate, e.g., reading from files
only if they open. Indeed, it is important to model the extent to
which the \emph{environment} determines the state after a command.
McBride~\cite{McBride11} observes that indexing input and output types
over the state effectively lets us specify interfaces in a
proof-relevant Hoare logic. Hancock and
Hyvernat~\cite{DBLP:journals/apal/HancockH06}
have explored the compositionality of indexed `interaction
structures', showing that it is possible to model both sharing and
independence of state between interfaces.

\paragraph{Session Types as Interface Indices}
Our \verb!pipe! is a simple implementation of processes
communicating according to a rather unsubtle protocol, with an
inevitable but realistic `broken pipe' failure mode. We should surely
aim for more sophisticated protocols and tighter compliance.
The interface for interaction on a channel should be indexed over
session state, ensuring that the requests arriving at a coordinating
multihandler match exactly.

\paragraph{Substructural Typing for Honesty with Efficiency}
Using \verb!Abort!, we know that the failed computation will not
resume under any circumstances, so it is operationally wasteful to
construct the continuation. Meanwhile, for \verb!State!, it is usual
for the handler to invoke the continuation \emph{exactly once},
meaning that there is no need to allocate space for the continuation
in the heap.  Moreover, if we want to make promises about the eventual
execution of operations, we may need to insist that handlers do invoke
continuations sooner or later, and if we want communicating systems to
follow a protocol, then they should not be free to drop or resend
messages. Linear, affine, and relevant type systems offer tools to
manage uses more tightly: we might profitably apply them to
continuations and the data structures in which they are stored.

\paragraph{Modules and Type Classes}
Frank's effect interfaces provide a form of modularity and abstraction,
tailored to effectful programming in direct style. It seems highly
desirable to establish the formal status of interfaces with respect to
other ways to deliver modularity, such as ML
modules~\cite{DBLP:conf/lfp/MacQueen84} and Haskell type
classes~\cite{DBLP:conf/popl/WadlerB89}.

\paragraph{Totality, Productivity, and Continuity}
At heart, Frank is a language for incremental transformation of
computation (command-response) trees whose node shapes are specified
by interfaces, but in the `background', whilst keeping the values
communicated in the foreground. Disciplines for \emph{total}
programming over treelike data, as foreground values, are the staple
of modern dependently typed programming languages, with the state of
the art continuing to advance~\cite{DBLP:conf/icfp/AbelP13}.  The
separation of client-like inductive structures and server-like
coinductive structures is essential to avoid deadlock (e.g., a server
hanging) and livelock (e.g., a client constantly interacting but
failing to return a value). Moreover, local \emph{continuity}
conditions quantifying the relationship between consumption and
production (e.g., \verb!spacer! consuming one input to produce two
outputs) play a key role in ensuring global termination or
productivity. Guarded recursion seems a promising way to capture these
more subtle requirements~\cite{DBLP:conf/icfp/AtkeyM13}.

Given that we have the means to negotiate purity locally whilst still
programming in direct style, it would seem a missed opportunity to
start from anything other than a not just \emph{pure} but \emph{total}
base. To do so, we need to refine our notion of `ability' with a
continuity discipline and check that programs obey it, deploying the
same techniques total languages use on foreground data for the
background computation trees. McBride has shown that general recursion
programming fits neatly in a Frank-like setting by treating recursive
calls as abstract commands, leaving the semantics of recursion for a
handler to determine~\cite{DBLP:conf/mpc/McBride15}.

\section{Conclusion}
\label{sec:con}
We have described our progress on the design and implementation of
Frank, a language for direct-style programming with locally managed
effects. Key to its design is the generalisation of function
application to operator application, where an operator is $n$-adic and
may handle effects performed by its arguments. Frank's effect type
system statically tracks the collection of permitted effects and
convenient syntactic sugar enables lightweight effect polymorphism in
which the programmer rarely needs to read or write any effect
variables.

It is our hope that Frank can be utilised as a tool for tackling the
programming problems we face in real life. Whether we are writing
elaborators for advanced programming languages, websites mediating
exercises for students, or multi-actor communicating systems, our
programming needs increasingly involve the kinds of interaction and
control structures which have previously been the preserve of
heavyweight operating systems development. It should rather be a joy.

%
%
%
%
%












\acks

We would like to thank the following people: Fred McBride for the idea
of generalising functions to richer notions of context;
Stevan Andjelkovic, Bob Atkey, James McKinna, Gabriel Scherer, Cameron Swords,
and Philip Wadler for helpful feedback;
Michael Adams and Adam Gundry for answering questions regarding their
respective works and for providing source code used as inspiration;
and Daniel Hillerstr\"om for guidance on OCaml Multicore.
This work was supported by EPSRC grants EP/J014591/1, EP/K034413/1,
and EP/M016951/1, a Royal Society Summer Internship, and the
Laboratory for Foundations of Computer Science.



\bibliographystyle{abbrvnat}
\bibliography{frankly}






\end{document}